\documentclass[twocolumn,showpacs,preprintnumbers,amsmath,amssymb,superscriptaddress]{revtex4-1}

\usepackage[utf8]{inputenc}
\usepackage[T1]{fontenc}
\usepackage{ulem}
\usepackage{tabularx}

\usepackage{amsmath}
\usepackage{mathtools}
\usepackage{braket}
\usepackage{tensor}
\usepackage{xcolor}
\usepackage[abbreviations]{siunitx}
\usepackage{booktabs}
\usepackage{float}

\usepackage{graphicx,color,rotating,pifont}
\usepackage{siunitx}
\usepackage{bm}
\usepackage{ae}
\usepackage{dcolumn}
\usepackage{txfonts}
\usepackage{tensor}
\usepackage{booktabs}
\usepackage{xspace}
\usepackage{mathrsfs}
\usepackage{amssymb} 
\usepackage{amsmath}
\usepackage{esvect}
\usepackage{diagbox}
\usepackage{soul}
\setstcolor{red}
\setul{}{.2ex}

\usepackage{hyperref}
\hypersetup{
	colorlinks=true, 
	linktoc=all,     
	linkcolor=blue,  
}

\usepackage{cleveref}

\renewcommand{\vec}[1]{\ensuremath{\bm{#1}}}

\newcommand{\beq}{\begin{equation}}
\newcommand{\eeq}{\end{equation}}
\newcommand{\beqn}{\begin{eqnarray}}
\newcommand{\eeqn}{\end{eqnarray}}
\newcommand{\bsub}{\begin{subequations}}
\newcommand{\esub}{\end{subequations}}
\newcommand{\bpm}{\begin{pmatrix}}
\newcommand{\epm}{\end{pmatrix}}

\DeclareSIUnit{\fm}{\femto\meter}
\DeclareSIUnit{\MeVc}{\MeV\per\text{\ensuremath{c}}}

\allowdisplaybreaks[3]

\begin{document} 

\title{Statistical uncertainty quantification for multireference covariant density functional theory}
 
 \author{X. Zhang}  
 \affiliation{School of Physics and Astronomy, Sun Yat-sen University, Zhuhai 519082, P.R. China} 
    \affiliation{Department of Physics, Kyoto University, Kyoto 606-8502, Japan}    
  
   \author{C. C. Wang}  
  \affiliation{School of Physics and Astronomy, Sun Yat-sen University, Zhuhai 519082, P.R. China} 
\affiliation{Graduate School of China Academy of Engineering Physics, Beijing 100193, China}

   \author{C. R. Ding}  
  \affiliation{School of Physics and Astronomy, Sun Yat-sen University, Zhuhai 519082, P.R. China}

  \author{J. M. Yao}    
  \email{Contact author: yaojm8@sysu.edu.cn}
  \affiliation{School of Physics and Astronomy, Sun Yat-sen University, Zhuhai 519082, P.R. China}   
    \affiliation{Yukawa Institute for Theoretical Physics, Kyoto University, Kyoto, 606-8502, Japan}    
  
\begin{abstract} 
We present a theoretical framework to quantify statistical uncertainties in covariant density functional theory (CDFT) for both nuclear matter and finite nuclei, based on a relativistic point-coupling energy density functional (EDF). By sampling approximately one million parameter sets, with nine parameters varied around their values in the PC-PK1 functional, we construct a probability density function for nuclear matter properties. Incorporating empirical values of nuclear matter at saturation density and those of predictions from chiral nuclear forces, and measured  $B(E2)$ values of finite nuclei, we infer posterior distributions for the model parameters within a Bayesian framework. These posterior distributions are then propagated to the low-lying states of finite nuclei using the newly developed subspace-projected (SP)-CDFT approach, in which the wave functions of target EDF parameter sets are expanded in a subspace spanned by low-lying states obtained from a set of training parameterizations. We find that the observables of low-lying states in deformed nuclei \nuclide[150]{Nd} and \nuclide[150]{Sm} are well reproduced once statistical uncertainties are taken into account. In contrast, those of near spherical nuclei \nuclide[136]{Xe} and \nuclide[136]{Ba} remain difficult to describe within the present framework, a limitation that is expected to be alleviated by extending the model space to include quasiparticle excitations.

\end{abstract}
  
\maketitle

\section{Introduction}

Nuclear density functional theory (DFT) provides a microscopic and self-consistent framework for a unified description of finite nuclei and neutron-star matter based on a universal energy density functional (EDF)~\cite{Bender:2003RMP,Vretenar:2005PR,Meng:2005PPNP,Drut:2010PPNP}. In its simplest implementation, namely the self-consistent mean-field approximation, the complex nuclear many-body problem is effectively reduced to an equivalent one-body problem. The energy of a nuclear system is then approximated as a functional of the powers and gradients of nuclear densities and currents, which according to the Kohn-Sham scheme~\cite{Kohn:1965}, can be represented in terms of auxiliary single-particle wave functions. This approach, known as single-reference (SR)-DFT, has achieved a great success in reproducing the properties of nuclear matter around saturation density  and the ground states of finite nuclei across the entire nuclear chart~\cite{ Lalazissis:1999,Geng:2005,Erler:2012,Erler:2012,Agbemava:2019PRC,DRHBcMassTable:2024}.
Despite its success, nuclear DFT faces significant challenges due to discrepancies in predictions made by different EDFs. These discrepancies lead to considerable uncertainties, particularly  for the equation of state of nuclear matter at densities far from the saturation point~\cite{Li:2008PR}, and for neutron-rich nuclei with limited experimental data~\cite{Afanasjev:2016PRC_uncertainty}. In this context, it is necessary to quantify the errors of nuclear DFT which consist of systematic and statistical components~\cite{Dobaczewski:2014}. Since the currently widely used nuclear EDFs are derived from phenomenological nucleon-nucleon effective interactions—such as non-relativistic Skyrme~\cite{Skyrme:1959,Vautherin:1972} and Gogny~\cite{Gogny:1970,Decharge:1980} forces, as well as relativistic covariant EDFs~\cite{Walecka:1974,Reinhard:1989,Ring:1996,Meng:2016_CDFT}—it is difficult to quantify the systematic errors of nuclear DFT. This remains the case despite efforts to develop new generations of EDFs~\cite{Schwenk:2004,Gambacurta:2011,Grasso:2016,Yang:2017,Bonnard:2018,Liang:2018,Burrello:2021,Marino:2021,NavarroPerez:2018PRC,Zurek:2024PRC} inspired by effective field theories or based on ab initio calculations, as reviewed in~\cite{Drut:2010PPNP,Shen:2019,Grasso:2019PPNP,Furnstahl:2020EPJA}. In contrast, the statistical error associated with a given EDF, which arises from variations in the parameters around their optimal values, can be quantified using statistical methods. Over the past decade, significant progress has been made in quantifying the statistical uncertainties of DFT predictions for nuclear ground-state bulk properties~\cite{Goriely:2014,Dobaczewski:2014,McDonnell:2015PRL,Agbemava:2014PRC,Agbemava:2019PRC}, and in identifying potential correlations between nuclear matter properties~\cite{Giuliani:2022} and neutron-star observables~\cite{Salinas:2023PRC,Sun:2023}.

Extending DFT to study energy spectra and transition strengths of nuclear low-lying states typically requires going beyond the mean-field approximation. In the SR-DFT, nuclear wave function is approximated as a product of auxiliary single-particle wave functions determined with the variational principle. This approach ensures the solution corresponds to a local energy minimum within the restricted Hilbert space, but it does not preserve the symmetry structure of nuclear many-body Hamiltonians. A common example is the introduction of deformation and pairing correlations in the SR-DFT for open-shell nuclei, which violate the $S(O3)$ and $U(1)$ symmetries. As a result, the quantum numbers associated with angular momentum and particle number are missing in nuclear wave functions—critical for studies of nuclear low-lying spectroscopy~\cite{Yao:2023HandBook}. The restoration of broken symmetries and the inclusion of dynamical correlations from fluctuations around the equilibrium shape in the SR-DFT can be achieved through quantum-number projection and the generator coordinate method (GCM)\cite{Ring:1980,Yao:2023HandBook}. This extended framework, known as multireference DFT (MR-DFT), has been successfully applied to study nuclear low-lying states~\cite{Bender:2003RMP,Bender:2008,Rodriguez:2010PRC,Yao:2010,Niksic:2011PPNP,Robledo:2018JPG,Sheikh:2021qv,SunXX:2021_Proj,Zhou:2023IJMPE}, as well as the nuclear matrix elements (NMEs) of $0\nu\beta\beta$ decay~\cite{Rodriguez:2010PRL,Song:2014,Yao:2022PPNP,Wang:2024PRC}. 

With advances in nuclear technology and methodologies, nuclear physics is entering an era of high precision. Accurate measurements of atomic and nuclear spectroscopy in neutron-rich nuclei~\cite{Chupp:2019RMP,Arrowsmith-Kron:2024}, as well as the half-lives of rare nuclear processes~\cite{Severijns:2006RMP,Snowmass_EDM:2022,Snowmass_NLDBD:2022}, demand precise modeling of nuclear low-lying states and the corresponding NMEs. Therefore, quantifying the theoretical uncertainties of these physical quantities is essential for making meaningful comparisons with other models and available data. However, the uncertainties in nuclear low-lying states have been scarcely studied within EDF frameworks, primarily due to the computational intensity required for enormous repeated calculations with varying EDF parameter sets. 


Recently, we performed a Bayesian analysis of nuclear low-lying states and $0\nu\beta\beta$ decay within a covariant EDF framework, enabled by the newly developed subspace-projected covariant density functional theory (SP-CDFT)~\cite{Zhang:2024_Letter}. This approach combines multireference CDFT (MR-CDFT) with the eigenvector continuation (EC) method. The central idea of EC is to represent the eigenvector of a target Hamiltonian within a low-dimensional subspace spanned by the eigenvectors of a set of sampling Hamiltonians~\cite{Frame:2018PRL}. The efficiency and accuracy of EC, when coupled with various many-body methods, have been demonstrated in a wide range of toy models~\cite{Franzke:2022PLB,Sowinski:2022PRC,Baran:2023PRB,Franzke:2024PRC,Luo:2024PRC} as well as in nuclear structure and reaction studies~\cite{Ekstrom:2019PRL,Konig:2020fn,Furnstahl:2020PLB,Demol:2020PRC,Drischler:2021PLB,Bai:2021PRC,Sarkar:2021}; see also the reviews~\cite{Drischler:2022,Duguet:2023RMP}.
In the present work, we provide a detailed description of this framework and apply it to quantify statistical uncertainties in both nuclear matter properties and low-lying nuclear states. The analysis is carried out for deformed nuclei, \nuclide[150]{Nd} and \nuclide[150]{Sm}, as well as near-spherical systems, including \nuclide[136]{Xe} and \nuclide[136]{Ba}, within a Bayesian framework.
Comparing the predictions with their associated statistical uncertainties provides valuable insight into the strengths and limitations of the current implementation of MR-CDFT.

The remainder of this paper is organized as follows. In Sec.~\ref{sec:methods}, we introduce the theoretical framework, including SR-CDFT, MR-CDFT, and SP-CDFT. Section~\ref{sec:results} presents benchmark calculations and the quantification of statistical uncertainties for nuclear matter and low-lying nuclear states. Finally, Sec.~\ref{sec:summary} summarizes our findings and outlines future perspectives.

\section{The methods}
\label{sec:methods}
In this section, we present a self-contained description of the theoretical framework, including SR-CDFT~\cite{Vretenar:2005PR,Meng:2005PPNP,Meng:2016_CDFT} and MR-CDFT~\cite{Yao:2009PRC,Yao:2010,Niksic:2011PPNP,Zhou:2024PRC}, together with a more comprehensive introduction to SP-CDFT for nuclear low-lying states.  

\subsection{The CDFT for finite nuclei} 
In the SR-CDFT for finite nuclei, the nuclear wave function $\ket{\Phi(\mathbf{q},\mathbf{C})}$ is approximated as a Slater determinant, which is a product of the wave functions  $\psi_k$ of single-particle states. These single-particle states are determined by minimizing the following covariant EDF~\cite{Burvenich:2002PRC,Zhao:2010PRC},
 \beq
 \label{eq:Energy}
  E[\tau, \rho,\nabla\rho; \mathbf{C}] 
    =\int d^3r \Big[\tau(\boldsymbol{r})
    +\mathcal{E}^{\text {em}}(\boldsymbol{r})
    + \mathcal{E}^{\rm int}(\bm{r}) \Big].
\eeq%
The first term represents the kinetic energy of nucleons,
 \beq
 \label{kinetic}
 \tau(\bm{r})=
  \sum_k\,\varv_k^2~{\psi^\dagger_k (\bm{r})
    \left(\bm{\alpha}\cdot\bm{p} + \beta M - M\right )\psi_k(\bm{r})},
\eeq
where $\varv^2_k\in[0, 1]$ is the occupation probability of the $k$-th single-particle state, and its value is determined by the Bardeen–Cooper–Schrieffer (BCS) theory based on a zero-range pairing force~\cite{Yao:2009PRC}. The pairing strengths are taken as the values from Ref.~\cite{Song:2014}. The $M$ is the nucleon bare mass. The $\psi_k$ is a Dirac spinor,  and $\bm{\alpha}, \beta$ are Dirac matrices. Using the equation of motion for the static electromagnetic field $A_\mu(\bm{r})$, one finds the second term in \eqref{eq:Energy} for the energy of electromagnetic interaction between protons,
 \beq
 \mathcal{E}^{\rm em}(\bm{r})
 =\frac{e}{2} A_\mu(\bm{r}) j^{\mu}_{V, p}(\bm{r}) 
\eeq
where $j^{\mu}_{V, p}(\bm{r})$ is the proton current in coordinate space, and $e$ is the charge of bare proton. The last term in \eqref{eq:Energy} is for the energy of nucleon-nucleon effective interactions,
 \beqn
 \label{eq:Int_EDF}
 \mathcal{E}^{\rm int}(\bm{r})
 &=& \frac{\alpha_S}{2}\rho_S^2+\frac{\beta_S}{3}\rho_S^3
    + \frac{\gamma_S}{4}\rho_S^4+\frac{\delta_S}{2}\rho_S\triangle \rho_S\nonumber \\
 &&+ \frac{\alpha_V}{2}j_\mu j^\mu + \frac{\gamma_V}{4}(j_\mu j^\mu)^2 +
       \frac{\delta_V}{2}j_\mu\triangle j^\mu  \nonumber\\
 && +  \frac{\alpha_{TV}}{2}\vec j^{\mu}_{TV}\cdot(\vec j_{TV})_\mu+\frac{\delta_{TV}}{2}
    \vec j^\mu_{TV}\cdot\triangle(\vec j_{TV})_{\mu}\nonumber\\
    &\equiv &  \sum^9_{\ell=1} c_\ell  \mathcal{E}^{NN}_\ell(\boldsymbol{r}),
\eeqn
which is decomposed into nine terms. Each term comes with a low-energy coupling constant (LEC) $c_\ell$. All the nine LECs are collectively labeled as $\mathbf{C}=\{\alpha_S, \beta_S, \gamma_S, \delta_S, \alpha_V, \gamma_V, \delta_V, \alpha_{TV}, \delta_{TV}\}$. The subscripts $(S, V)$ indicate the scalar and vector types of coupling vertices in Minkowski space, respectively, and $T$ for the vector in isospin space.  The symbols $\alpha_S$, $\alpha_V$,   and $\alpha_{TV}$ denote the coupling constants associated with four-fermion contact interaction terms, while $\beta_S$, $\gamma_S$, and $\gamma_V$ represent nonlinear self-interaction terms. Additionally, $\delta_S$, $\delta_V$,  and $\delta_{TV}$ denote the coupling constants for gradient terms to simulate finite-range effects of nuclear force.  
 
 It is seen from \eqref{eq:Int_EDF} that the interaction energy is a functional of the local scalar density $\rho_S(\bm{r})$, four-component currents $j^\mu_{V}(\bm{r}), \vec j^\mu_{TV}(\bm{r})$ and their derivatives, where the density and currents are determined by the single-particle wave functions,
\bsub%
\label{currents}
\beqn%
  \label{E13a}
  \rho_S(\bm{r})          &=&\sum_{k }\varv^2_k\bar\psi_k(\bm{r})\psi_k(\bm{r}),\\
  \label{E13b}
  j^\mu_{V}(\bm{r})       &=&\sum_{k }\varv^2_k\bar\psi_k(\bm{r})\gamma^\mu\psi_k(\bm{r}),\\
  \label{E13c}
  \vec j^\mu_{TV}(\bm{r}) &=&\sum_{k }\varv^2_k\bar\psi_k(\bm{r})\vec\tau\gamma^\mu\psi_k(\bm{r}).
\eeqn%
\esub%
Here, $\vec{\tau}$ indicates an vector in the isospin space. The $k$ index runs over all single-particle states under the {\it no-sea} approximation~\cite{Ring:1996}.   Minimization of the EDF in (\ref{eq:Energy}) with respect to $\bar\psi_k$
 gives rise to the Dirac equation for the single nucleons
 \beqn
  \label{eq:DiracEq}
  [\gamma_\mu(i\partial^\mu-V^\mu)-(M+\Sigma_S)]\psi_k(\bm{r})=0.
 \eeqn
 The single-particle effective Hamiltonian contains scalar $\Sigma_S(\bm{r})$ and vector
 $V^\mu(\bm{r})$ potentials
 \beq
 \label{eq:potential} 
   V^\mu(\bm{r})=\Sigma^\mu+\vec\tau\cdot\vec\Sigma^\mu_{TV},
 \eeq
 where  
 \bsub\beqn
  \Sigma_S           &=&\alpha_S\rho_S+\beta_S\rho^2_S+\gamma_S\rho^3_S+\delta_S\triangle\rho_S,\\
  \Sigma^\mu   &=&\alpha_Vj^\mu_V +\gamma_V (j^\mu_V)^3
                       +\delta_V\triangle j^\mu_V + e A^\mu,\\
  \vec\Sigma^\mu_{TV}&=& \alpha_{TV}\vec j^\mu_{TV}+\delta_{TV}\triangle\vec j^\mu_{TV}.
 \eeqn\esub
In order to generate nuclear mean-field wave functions  $\ket{\Phi(\mathbf{q},\mathbf{C})}$ with different deformation parameters $\mathbf{q}$, we  impose a  quadrupole constraint on  the mass quadrupole moment during the above minimization procedure~\cite{Yao:2009PRC,Ring:1980}. In this work, only axially-deformed parity-conserving mean-field states are considered, in which case the symbol $\mathbf{q}$ is simply the quadrupole deformation parameter $\beta_{20}$ determined by 
\beq 
\beta_{20}=\frac{4\pi}{3AR^2}\bra{\Phi(\mathbf{q},\mathbf{C})}\hat Q_{20}\ket{\Phi(\mathbf{q},\mathbf{C})},
\eeq 
where $R=1.2A^{1/3}$ fm with $A$ being nuclear mass number. The quadrupole moment operator is defined as $\hat Q_{20}  = r^2Y_{20}$, where $Y_{20}$ is the rank-2 spherical harmonic function.

\subsection{The MR-CDFT for nuclear low-lying states} 
In the MR-CDFT, wave function of nuclear low-lying state is constructed as a superposition of quantum-number projected mean-field wave functions~\cite{Ring:1980},
\begin{equation}
  \ket{\Psi^{JNZ}_\nu(\mathbf{C})}=\sum^{N_\mathbf{q}}_{\mathbf{q}} f^{JNZ}_{\nu}(\mathbf{q}, \mathbf{C}) \ket{JNZ; \mathbf{q}, \mathbf{C}},
\end{equation}
 where $\nu$ distinguishes different states with the same quantum numbers $JM$. The basis function is constructed as
 \begin{equation}
   \ket{JNZ; \mathbf{q}, \mathbf{C}} \equiv  \hat P^J_{M0} \hat P^N\hat P^Z\vert \Phi(\mathbf{q}, \mathbf{C})\rangle,
 \end{equation}
 with $\hat P^{J}_{M0}$ and  $\hat{P}^{N, Z}$ are the projection operators that extract the component with the angular momentum $J$ and its $z$-component $K=0$, neutron number $N$, proton number $Z$, 
\begin{subequations}
\begin{align}
    \hat P^{J}_{MK}&=\dfrac{2J+1}{8\pi^2}\int d\Omega D^{J\ast}_{MK}(\Omega) \hat R(\Omega),\\
    \hat P^{N_\tau} &= \dfrac{1}{2\pi}\int^{2\pi}_0 d\varphi_{\tau}  e^{i\varphi_{\tau}(\hat N_\tau-N_\tau)},
\end{align} 
\end{subequations} 
where   $D^{J\ast}_{MK}(\Omega)$ is the Wigner-D function of the Euler angles $\Omega$. The mean-field wave functions $\ket{\Phi(\mathbf{q},\mathbf{C})}$ are generated from the above self-consistent CDFT  calculation~\cite{Yao:2009PRC}. The weight function $f^{JNZ}_{\nu}(\mathbf{q}, \mathbf{C})$ is determined with the variational principle which leads to the Hill-Wheeler-Griffin (HWG) equation~\cite{Hill:1953,Ring:1980},
\begin{eqnarray}
\label{eq:HWG}
\sum_{\mathbf{q}'}
\Bigg[{\cal H}^{\mathbf{C}}(\mathbf{q}, \mathbf{q}')
-E_{\nu, \mathbf{C}}^{JNZ}{\cal N}^{\mathbf{C}}(\mathbf{q}, \mathbf{q}') \Bigg] f^{JNZ}_{\nu}(\mathbf{q}', \mathbf{C})=0,
\end{eqnarray}
where the Hamiltonian kernel and norm kernel are defined by
\bsub
\label{eq:GCM_kernel}
\beqn 
      {\cal N}^{\mathbf{C}}(\mathbf{q}, \mathbf{q}')
     &=& \bra{JNZ; \mathbf{q}, \mathbf{C}} JNZ; \mathbf{q}',\mathbf{C}\rangle,\\
     {\cal H}^{\mathbf{C}}(\mathbf{q}, \mathbf{q}')
    &=& \bra{JNZ; \mathbf{q}, \mathbf{C}} \hat H(\mathbf{C})\ket{JNZ; \mathbf{q}',\mathbf{C}}.
\eeqn
\esub 
 The Hamiltonian kernels ${\cal H}^{\mathbf{C}}(\mathbf{q}, \mathbf{q}')$ are evaluated with the generalized Wick theorem~\cite{Balian:1969}. In particular, the energy overlap is determined with the mixed-density prescription~\cite{Sheikh:2021qv,Yao:2022PPNP}.

 The electric quadrupole ($E2$) transition strength for $J^\pi_{i, \nu_{i}} \rightarrow J^\pi_{f, \nu_{f}}$ from the MR-CDFT calculation of a given parameter set $\mathbf{C}$ of EDF is determined by
  \begin{eqnarray}
\label{eq:BE2} 
&&B^\mathbf{C}(E2; J^\pi_{i, \nu_{i}} \rightarrow J^\pi_{f, \nu_{f}})  \nonumber \\
&=&\frac{1}{2 J_{i}+1}\left|\sum_{\mathbf{q}^{\prime}, \mathbf{q}} 
f^{J_fNZ}_{\nu_f}(\mathbf{q}', \mathbf{C})f^{J_iNZ}_{\nu_i}(\mathbf{q}, \mathbf{C})\right.\nonumber\\
&&\left.\times
\bra{J_fNZ; \mathbf{q}', \mathbf{C}}|\hat{Q}^{(e)}_{2}|\ket{J_iNZ;\mathbf{q}, \mathbf{C}} \right|^{2},
 \end{eqnarray} 
where the reduced matrix element  is determined as 
  \begin{eqnarray}
&&\bra{J_fNZ; \mathbf{q}', \mathbf{C}}|\hat{Q}^{(e)}_{2}|\ket{J_iNZ;\mathbf{q}, \mathbf{C}}  \nonumber\\
&=& \left(2 J_{f}+1\right)  (-1)^{J_{f}}\sum^2_{\mu=-2}\left(\begin{array}{ccc}
J_{f} & 2 & J_{i} \\
0 & \mu & -\mu
\end{array}\right)\nonumber\\
&&\times \bra{\Phi(\mathbf{q}^{\prime}, \mathbf{C})} er^2Y_{2\mu}  \hat P^{J_i}_{-\mu0} \hat{P}^{N} \hat{P}^{Z}\ket{\Phi(\mathbf{q}, \mathbf{C})}. 
 \end{eqnarray}  

 It is noted that for each parameter set of EDF, one needs to evaluate about $N^2_{\mathbf{q}}$ kernels, the calculation of which is usually very time consuming.  The computational cost grows rapidly with the number of mesh points in the projection operators. Thus, it has been challenging to quantify the statistical uncertainty of the MR-CDFT study for nuclear low-lying states as it requires massive repetitive calculations based on different EDF parameter sets. 

\subsection{Emulating MR-CDFT with the SP-CDFT}
 In this subsection, we introduce the SP-CDFT($N_t, k_{\rm max}$) as an emulator of the MR-CDFT for nuclear low-lying states, based on the EC method.  The wave function $\ket{\Psi^{JNZ}_k(\mathbf{C}_\odot)}$ of the $k$-th state for a target EDF labeled with $\mathbf{C}_\odot]$ is constructed as a superposition of the wave functions $\ket{\Psi^{JNZ}_\nu(\mathbf{C}_t)}$ of the first $k_{\rm max}$ states, 
\begin{equation}
\label{eq:EC_GCM_wfs}
   \ket{\bar\Psi^{JNZ}_k(\mathbf{C}_\odot)}
   =\sum^{k_{\rm max}}_{\nu=1}\sum^{N_t}_{t=1} \bar f^{JNZ}_{k, \mathbf{C}_\odot}(\nu, \mathbf{C}_t)\ket{\Psi^{JNZ}_\nu(\mathbf{C}_t)},
\end{equation}
where $k\in [1,2,\cdots, k_{\rm max}]$.  The mixing coefficient  $\bar f^{JNZ}_{k, \mathbf{C}_\odot}(\nu, \mathbf{C}_t)$ is determined by the following equation,
\beq 
 \sum^{k_{\rm max}}_{\nu'=1}\sum^{N_t}_{t'=1} \Bigg[ 
 \mathscr{H}^{\nu\nu'}_{tt'}(\mathbf{C}_\odot) 
 - \bar E_{k, \mathbf{C}_\odot}^{JNZ}  \mathscr{N}^{\nu, \nu'}_{tt'} \Bigg]  \bar f^{JNZ}_{k, \mathbf{C}_\odot}(\nu', \mathbf{C}_{t'})=0,
\eeq 
Here, we define the norm and Hamiltonian kernels of the EC method for a target EDF $E[\rho, \nabla\rho; \mathbf{C}_\odot]$ as below
\bsub
\label{eq:EC_kernel}
\beqn 
      \mathscr{N}^{\nu\nu'}_{tt'} 
     &=& \bra{\Psi^{JNZ}_\nu(\mathbf{C}_t)}\Psi^{JNZ}_{\nu'}(\mathbf{C}_{t'}\rangle,\\
   \mathscr{H}^{\nu\nu'}_{tt'}(\mathbf{C}_\odot) 
    &=& \bra{\Psi^{JNZ}_{\nu}(\mathbf{C}_t)} \hat H(\mathbf{C}_\odot) \ket{\Psi^{JNZ}_{\nu'}(\mathbf{C}_{t'})}.
\eeqn
\esub
 The main ingredients  of the SP-CDFT are the norm kernels,
\beqn 
 \mathscr{N}^{\nu\nu'}_{tt'} 
 &=& \bra{\Psi^{JNZ}_\nu(\mathbf{C}_t)}\Psi^{JNZ}_{\nu'}(\mathbf{C}_{t'})\rangle\nonumber\\
 &=&\sum_{\mathbf{q}, \mathbf{q}'}f^{JNZ}_\nu(\mathbf{q}, \mathbf{C}_t) f^{JNZ}_{\nu'}(\mathbf{q}', \mathbf{C}_{t'}) \nonumber\\ 
&&\times  \bra{JNZ; \mathbf{q}, \mathbf{C}_t} JNZ; \mathbf{q}',\mathbf{C}_{t'}\rangle, 
\label{wq:EC_norm_kernel}
\eeqn 
and Hamiltonian kernels, which can be efficiently determined as follows,
\beqn 
    \mathscr{H}^{\nu\nu'}_{tt'}(\mathbf{C}_\odot) 
    &=&  \bra{\Psi^{JNZ}_\nu(\mathbf{C}_t)} \hat H(\mathbf{C}_\odot) \ket{\Psi^{JNZ}_{\nu'}(\mathbf{C}_{t'})} \nonumber\\
    &=&\sum_{\mathbf{q}, \mathbf{q}'}f^{JNZ}_\nu(\mathbf{q}, \mathbf{C}_t) f^{JNZ}_{\nu'}(\mathbf{q}', \mathbf{C}_{t'}) \nonumber\\ 
&&\times \bra{JNZ; \mathbf{q}, \mathbf{C}_t} \hat H(\mathbf{C}_\odot)\ket {JNZ; \mathbf{q}',\mathbf{C}_{t'}}.
\label{wq:EC_Hamiltonian_kernel}
\eeqn 
For the configurations with $K=0$, the configuration-dependent Hamiltonian kernel is simplified as 
\beqn 
 &&\bra{JNZ; \mathbf{q}, \mathbf{C}_t} \hat H(\mathbf{C}_\odot)\ket {JNZ; \mathbf{q}',\mathbf{C}_{t'}} \nonumber \\
 &=& \bra{\Phi(\mathbf{q},\mathbf{C}_t)} \hat H(\mathbf{C}_\odot)
 \hat P^J_{00}\hat P^N\hat P^Z \ket{\Phi(\mathbf{q}',\mathbf{C}_{t'})} \nonumber \\
 &=& \frac{2J+1}{2}\int d^{J}_{00}(\cos\theta) d(\cos\theta) \int \frac{e^{-iN\varphi_n}}{2\pi} d\varphi_n 
 \int \frac{e^{-iN\varphi_p}}{2\pi} d\varphi_p \nonumber \\
 &&\times \bra{\Phi(\mathbf{q},\mathbf{C}_t)} \hat H(\mathbf{C}_\odot)
  e^{i\theta\hat J_y}e^{i\varphi_n\hat N}e^{i\varphi_p\hat Z}\ket{\Phi(\mathbf{q}',\mathbf{C}_{t'})},
  \eeqn
where the energy overlap is evaluated with the mixed-density prescription,
\beqn 
 &&\frac{\bra{\Phi(\mathbf{q},\mathbf{C}_t)} \hat H(\mathbf{C}_\odot)
  e^{i\theta\hat J_y}e^{i\varphi_n\hat N}e^{i\varphi_p\hat Z}\ket{\Phi(\mathbf{q}',\mathbf{C}_{t'})}}
  {\bra{\Phi(\mathbf{q},\mathbf{C}_t)}  
  e^{i\theta\hat J_y}e^{i\varphi_n\hat N}e^{i\varphi_p\hat Z}\ket{\Phi(\mathbf{q}',\mathbf{C}_{t'})}}\nonumber\\
  &=&\int d^3r \Big[\tilde\tau(\boldsymbol{r})
    +\tilde{\mathcal{E}}^{\text {em}}(\boldsymbol{r})
    + \sum^9_{\ell=1} c^{\odot}_\ell  \tilde{\mathcal{E}}^{NN}_\ell(\boldsymbol{r}) \Big].
\eeqn 
All the three terms on the right hand side depend on the generate coordinates and training parameter sets, i.e., $\mathbf{q},\mathbf{q}'$, and $\mathbf{C}_t,  \mathbf{C}_{t'}$. Among the three terms, only the interaction energy term depends on the parameters $c^{\odot}_\ell$ of the target EDF,
\beqn 
\label{eq:EDF_mixed_density}
&&\sum^9_{\ell=1} c^{\odot}_\ell  \tilde{\mathcal{E}}^{NN}_\ell(
\mathbf{q},\mathbf{C}_t; \mathbf{q}',\mathbf{C}_{t'})\nonumber\\
&=& \frac{\alpha^{\odot}_S}{2}\Tilde{\rho}_S^2+\frac{\beta^{\odot}_S}{3}\Tilde{\rho}_S^3
    + \frac{\gamma^{\odot}_S}{4}\Tilde{\rho}_S^4+\frac{\delta^{\odot}_S}{2}\Tilde{\rho}_S\triangle \Tilde{\rho}_S\nonumber \\
 &&+ \frac{\alpha^{\odot}_V}{2}\Tilde{j}_\mu \Tilde{j}^\mu + \frac{\gamma^{\odot}_V}{4}(\Tilde{j}_\mu \Tilde{j}^\mu)^2 +
       \frac{\delta^{\odot}_V}{2}\Tilde{j}_\mu\triangle \Tilde{j}^\mu \nonumber \\
 && +  \frac{\alpha^{\odot}_{TV}}{2}\vec \Tilde{j}^{\mu}_{TV}\cdot(\vec \Tilde{j}_{TV})_\mu+\frac{\delta^{\odot}_{TV}}{2}
    \vec \Tilde{j}^\mu_{TV}\cdot\triangle(\vec \Tilde{j}_{TV})_{\mu},
\eeqn 
where $\tilde\rho, \tilde j^\mu_i$ are the mixed densities and currents whose expressions have been presented in Refs.~\cite{Niksic:2006,Yao:2009PRC}. They are evaluated using the mean-field wave functions of the training sets, and thus do not depend on the parameter set $\mathbf{C}_\odot$ of the target EDF. It enables us to speed up the calculation of the corresponding Hamiltonian kernels of target EDF. The configuration-dependent  Hamiltonian kernel can be divided into parameter-free and parameter-dependent terms,
\beqn 
\label{eq:kernel_decomposition}
 &&\bra{JNZ; \mathbf{q}, \mathbf{C}_t} \hat H(\mathbf{C}_\odot)\ket {JNZ; \mathbf{q}',\mathbf{C}_{t'}} \nonumber \\
 &=&  \bra{JNZ; \mathbf{q}, \mathbf{C}_t} \hat H_0 \ket {JNZ; \mathbf{q}',\mathbf{C}_{t'}} \nonumber\\
 &&
 + \sum^9_{\ell=1} f(c^{\odot}_\ell) \bra{JNZ; \mathbf{q}, \mathbf{C}_t} \hat H^{NN}_{\ell}(\mathbf{c^0_\ell})\ket {JNZ; \mathbf{q}',\mathbf{C}_{t'}} 
 \eeqn
 where the first term is given by the kinetic energy and electromagnetic energy, while the second term consists of nine $NN$ interaction terms.  During the sampling of parameter sets,   we introduce a scaling factor $f(c_\ell) = c_\ell / c^0_\ell$ for each parameter in $\mathbf{C}$, where $c^0_\ell$ is the value of the  optimized parameter set of the EDF, i.e., PC-PK1~\cite{Zhao:2010PRC} in this work.  The $\bra{JNZ; \mathbf{q}, \mathbf{C}_t} \hat H^{NN}_{\ell}(\mathbf{c^0_\ell)})\ket {JNZ; \mathbf{q}',\mathbf{C}_{t'}}$ are the $\ell$-th term in the Hamiltonian kernels with the optimized parameter set $\mathbf{C}_0$ sandwiched by the wave functions depending on the generate coordinates $\mathbf{q},\mathbf{q}'$ and training parameter sets $\mathbf{C}_t,  \mathbf{C}_{t'}$.
 This approach allows the parameters of the EDFs in the Hamiltonian kernel of the target EDF to be factorized out. As will be shown later, this method enables the execution of millions of SP-CDFT calculations for different parameter sets of EDFs, with several orders of magnitude less computational time than MR-CDFT.

  The  $E2$ transition strength of the transition  $J^\pi_{i, k_{i}} \rightarrow J^\pi_{f, k_{f}}$ for a target parameter set $\mathbf{C}_\odot$ is determined by 
  \begin{eqnarray}
\label{eq:BE2_EC} 
&&B^{\mathbf{C}_\odot}(E2; J^\pi_{i, k_{i}} \rightarrow J^\pi_{f, k_{f}})  \nonumber \\
&\equiv&\frac{1}{2 J_{i}+1}\left| \bra{\Psi^{J_fNZ}_{k_f}(\mathbf{C}_{\odot})}| \hat{Q}^{(e)}_{2}|\ket{\Psi^{J_iNZ}_{k_i}(\mathbf{C}_{\odot})}\right|^{2},
 \end{eqnarray} 
where  the reduced matrix elements among the states of training EDFs are given by
\beqn 
&&\bra{\Psi^{J_fNZ}_{k_f}(\mathbf{C}_{\odot})}| \hat{Q}^{(e)}_{2}|\ket{\Psi^{J_iNZ}_{k_i}(\mathbf{C}_{\odot})}
  \nonumber\\
&=&\sum_{t_i,t_f;\nu_i,\nu_f} \bar{f}^{J_fNZ}_{k_f, \mathbf{C}_\odot}(\nu_f, \mathbf{C}_{t_f})\bar{f}^{J_iNZ}_{k_i, \mathbf{C}_\odot}(\nu_i, \mathbf{C}_{t_i}) \nonumber\\
&& \times\sum_{\mathbf{q}^{\prime}, \mathbf{q}} 
f^{J_fNZ}_{\nu_f}(\mathbf{q}', \mathbf{C}_{t_f})f^{J_iNZ}_{\nu_i}(\mathbf{q}, \mathbf{C}{t_i}) \nonumber\\
&& \times
\bra{J_fNZ; \mathbf{q}', \mathbf{C}_{t_f}}|\hat{Q}^{(e)}_{2}|\ket{J_iNZ;\mathbf{q}, \mathbf{C}_{t_i}}.
\eeqn

\section{Results and discussion}
 \label{sec:results}

The Dirac equation \eqref{eq:DiracEq} for  neutrons and protons in finite nuclei is solved self-consistently by expanding the large and small components of the Dirac spinor $\psi_k$ in  a set of spherical  harmonic oscillator (HO) basis with $12$ major shells.  The oscillator frequency is
given by $\hbar\omega_{0}=41A^{-1/3}$ MeV. The Gaussian-Legendre quadrature is used for the integral over the Euler angle $\theta$ in the calculations of the norm and hamiltonian kernels in \eqref{eq:GCM_kernel}. We choose the number of mesh points for the Euler angle $\theta$ in the interval $[0,\pi]$ as $N_\theta=12$, and those for gauge angles $\varphi_{\tau}$ in the interval $[0,2\pi]$ as $N_\varphi=5$, which are found to be able to give convergent results. More details on the calculation of nuclear low-lying states and NME of $0\nu\beta\beta$ decay can be found in Refs.~\cite{Yao:2014,Yao:2015}.

\subsection{Benchmark of SP-CDFT calculations}

\begin{figure}
        \centering
        \includegraphics[width=0.8\columnwidth]{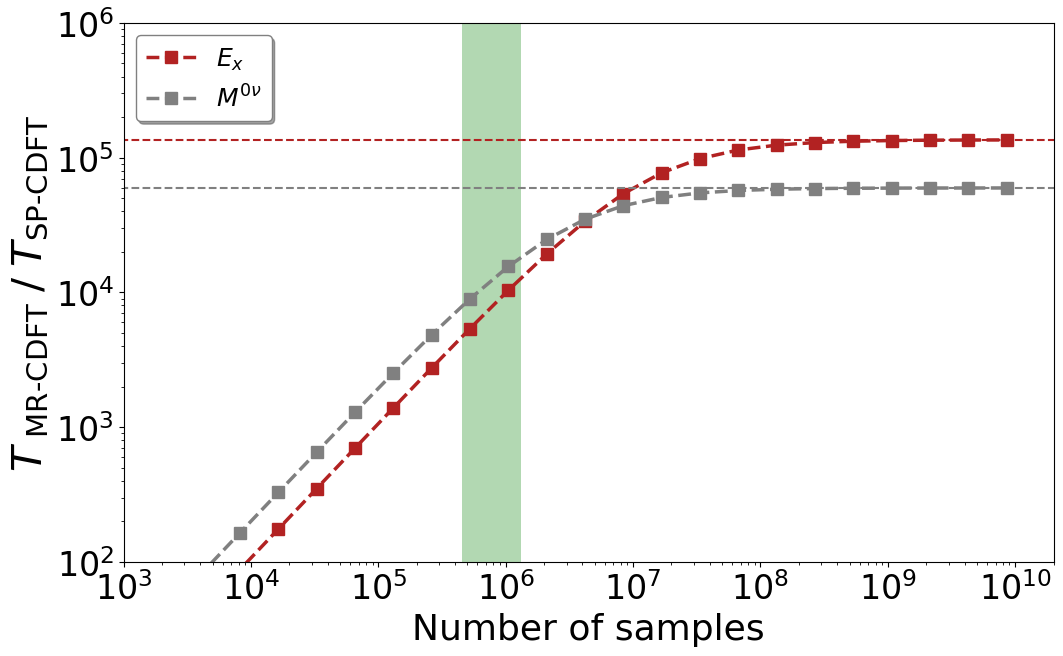}
    \caption{(Color online) The speed-up factor of SP-CDFT calculation for nuclear low-lying states and the NME $M^{0\nu}$ of $0\nu\beta\beta$ decay in \nuclide[150]{Nd}. The shaded area indicates the typical sample size.}
    \label{fig:time_complexity_comparison}
\end{figure}

\begin{figure}
    \centering    
    \includegraphics[width=0.8\columnwidth]{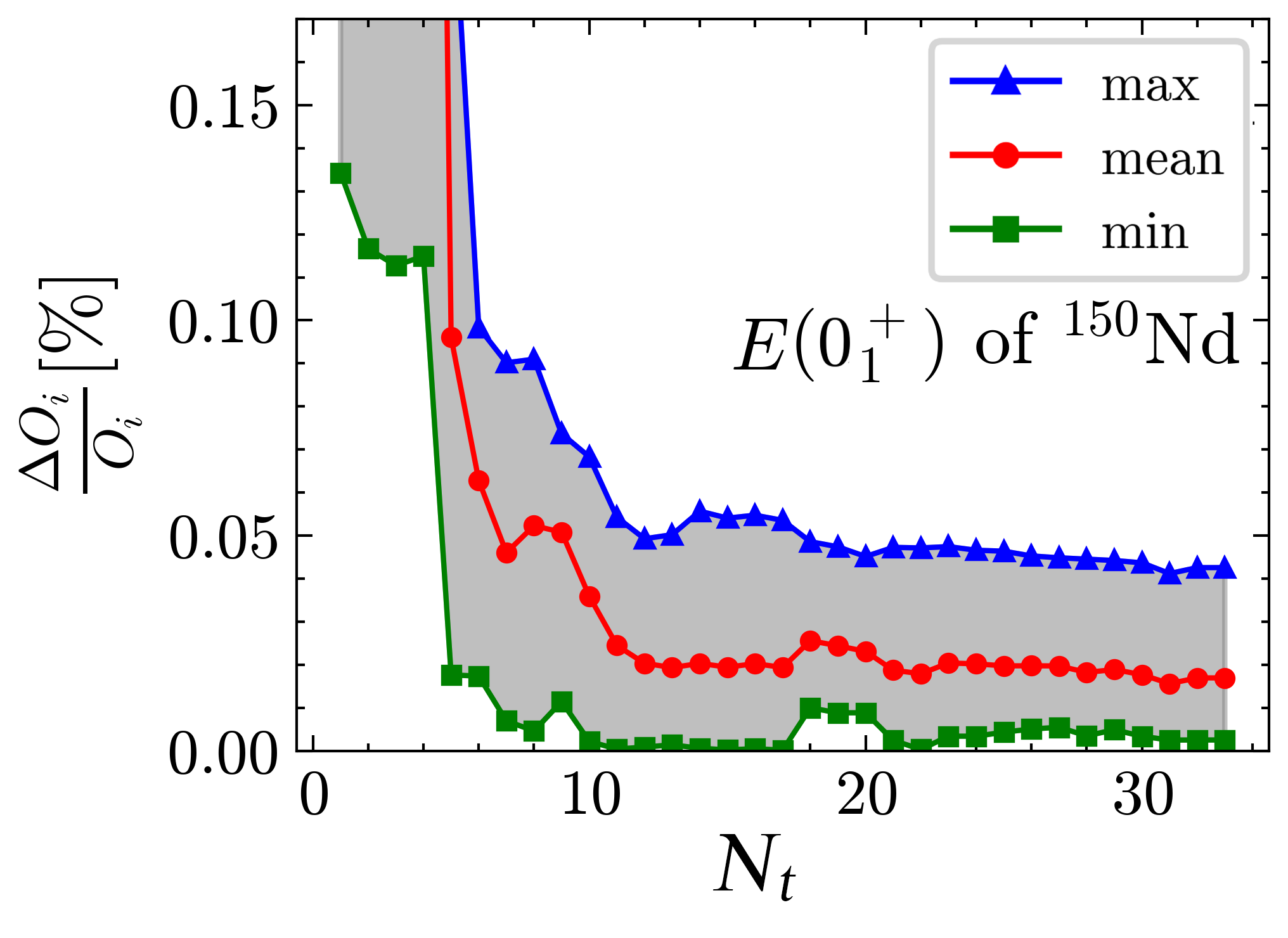}
    \caption{(Color online) The min-max and mean values of the relative errors of the ground-state energy of $^{150}$Nd from the SP-CDFT($N_t, 3$) calculations for the 64 testing sets as the function of the number $N_t$ of training sets. }
    \label{fig:convergence_test} 
\end{figure}

\begin{figure}
    \centering    \includegraphics[width=\columnwidth]{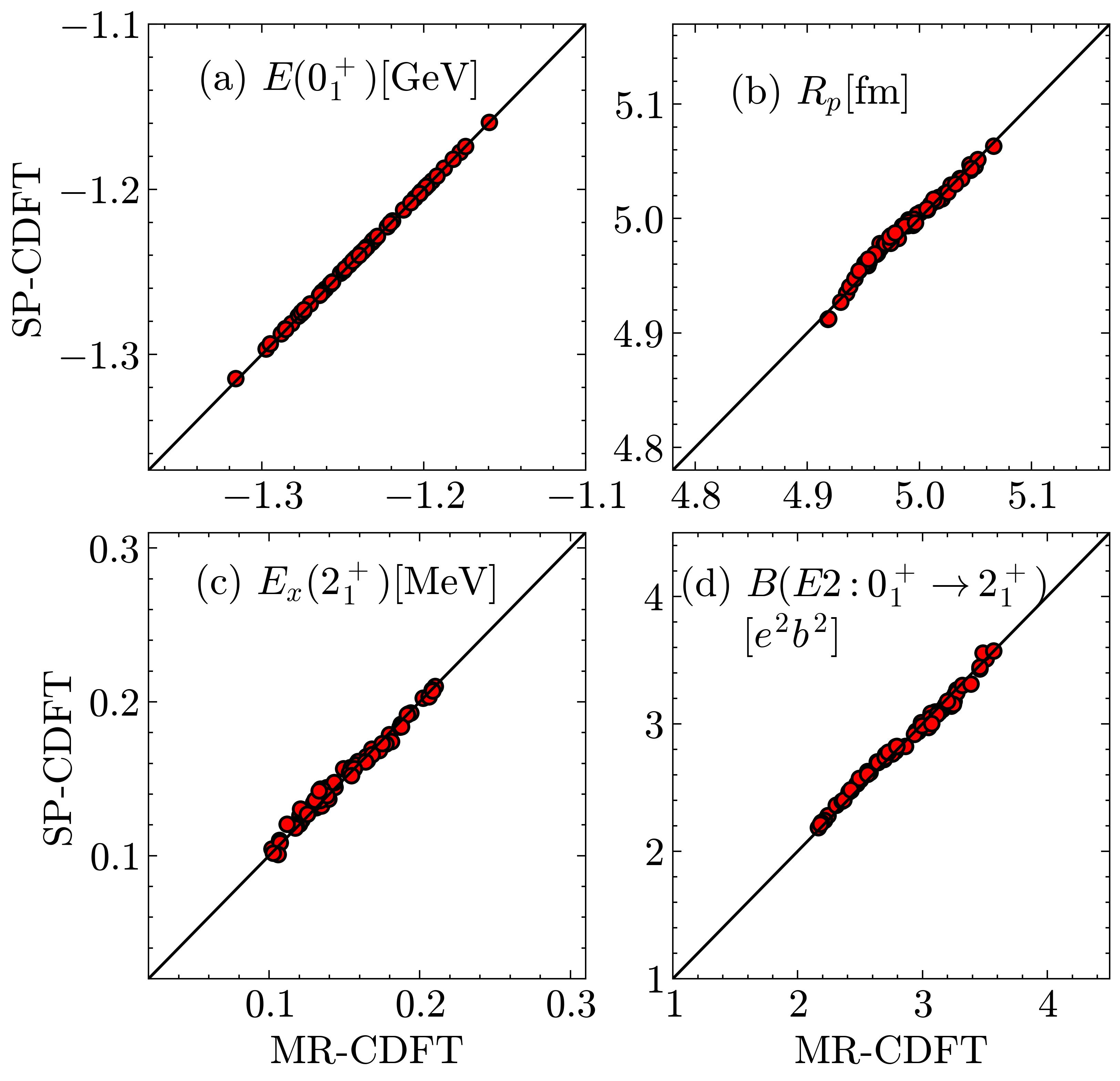}
    \caption{(Color online) Comparison of the ground-state properties  and low-lying states properties from the SP-CDFT(14,3) and MR-CDFT calculations for \nuclide[150]{Nd} based on 64 testing EDFs. }
    \label{fig:Nd150_bechmark} 
\end{figure}

The time complexity of MR-CDFT calculations for $N_{\mathbf{C}_\odot}$ target parameter sets is given by
\begin{align}
\label{eq:time_complexity_GCM}
    T_{\rm MR-CDFT} =O\Big(N^2_{\mathbf{q}}N_{\mathbf{C}_\odot}\Big)\Delta T_1,
\end{align}
where $\Delta T_1$ represents the computational time for each GCM kernel. In contrast,  the time complexity of the SP-CDFT($N_t$, $k_{\rm max}$) is composed of two parts,
\begin{align}
\label{eq:time_complexity_EC}
    T_{\rm SP-CDFT} = O\Big( N^2_{\mathbf{q}} N^2_t \Big)\Delta T_1 + O\Big(N^2_{\rm EC}N_{\mathbf{C}_\odot}\Big)\Delta T_2.
\end{align}
where the first term represents the computational time  of $N_t$ training sets, while the second term represents the time needed to evaluate the EC kernels of $N_{\mathbf{C}_\odot}$ target sets, $N_{\rm EC}=N_tk_{\rm max}$.
It is seen that $T_{\rm SP-CDFT}>T_{\rm MR-CDFT}$ for $N_\odot<N^2_t$. With the increase of $N_{\mathbf{C}_\odot}$, both $T_{\rm MR-CDFT}$ and $T_{\rm SP-CDFT}$ increase linearly, but with the slops of $N^2_{\mathbf{q}}\Delta T_1$ and $N^2_{\rm EC}\Delta T_2$, respectively. By breaking the interaction energy  (\ref{eq:Int_EDF}) into nine terms, one can compute the Hamiltonian kernels of EC efficiently using the GCM kernels of the training EDFs, see Eq.\eqref{eq:kernel_decomposition}. Quantitatively, we find $\Delta T_1/\Delta T_2\simeq 10^5$ for \nuclide[150]{Nd}. In other words, one would expect $T_{\rm SP-CDFT}\ll T_{\rm MR-CDFT}$ when the number of target sets $N_{\mathbf{C}_\odot}$  is sufficient large.  

Figure~\ref{fig:time_complexity_comparison} displays the speed-up factor, defined as the ratio of  $T_{\rm MR-CDFT}$  to $T_{\rm SP-CDFT}$ for nuclear low-lying states and the NME $M^{0\nu}$ of $0\nu\beta\beta$ decay, as a function of $N_{\mathbf{C}_\odot}$. The $M^{0\nu}$ is computed using the transition operators based on the standard mechanism, see Refs.~\cite{Song:2014,Yao:2015} for details.  Since the  $T_{\rm MR-CDFT}$ increases linearly with the number of samples, and the $T_{\rm SP-CDFT}$ is barely changing with it,  the speed-up factor $T_{\rm MR-CDFT}/T_{\rm SP-CDFT}$  increases almost linearly up to $10^4$ when the number of sampling EDFs reaches to $10^6$. It is also seen from Fig.~\ref{fig:time_complexity_comparison} that the speed-up factor reaches  asymptotically a limit as the
number of samples increases up to $10^8$, i.e., $T_{\rm MR-CDFT}/T_{\rm SP-CDFT}\to (N_{\mathbf{q}}/N_{\rm EC})^2 (\Delta T_1/\Delta T_2)$. A similar behavior has been found in Ref.~\cite{Konig:2020fn}.   In short, the SP-CDFT allows us, within half an hour using a PC,  to predict nuclear low-lying states for millions of EDFs which would otherwise take years with the MR-CDFT.

The accuracy of SP-CDFT($N_t, k_{\rm max}$) depends on the selected values of $N_t$ and $k_{\rm max}$. As demonstrated in Ref.\cite{Zhang:2024_Letter}, choosing $k_{\rm max} \geq 3$ effectively reproduces the excitation energy of the $2^+_1$ state. Figure~\ref{fig:convergence_test}  shows the relative error in the ground-state energy of \nuclide[150]{Nd} from SP-CDFT($N_t, k{\rm max}=3$) calculations across 64 test sets. Both the training sets and testing sets are sampled using the Latin hypercube sampling method \cite{Dutta:2020LHS}, which is commonly used to generate a representative sample of parameter values from a multidimensional distribution~\cite{McDonnell:2015PRL,Jiang2020,Sun:2024,Qiu:2024PLB}. Here, a uniform distribution is chosen as the probability distribution. The parameter ranges for Latin hypercube sampling are selected based on the uncorrelated tolerance of parameters with $\chi^2 \leq \chi^2_{\rm min} + 1$. As shown in Fig.~\ref{fig:convergence_test}, as $N_t$ increases to 14, the mean relative error decreases to 0.04\% and stabilizes. Consequently, we select $N_t = 14$ for the subsequent calculations.

Figure~\ref{fig:Nd150_bechmark}  compares the ground-state energies, root-mean-square (rms) proton radii, excitation energies $E_x(2^+_1)$, and $B(E2: 0^+_1 \to 2^+_1)$ values obtained from SP-CDFT ($14, 3$) and MR-CDFT calculations for \nuclide[150]{Nd} using 64 test sets. The data points align along the diagonal line, indicating strong agreement between the two methods. To quantify the accuracy of the emulator, the standard deviation of the SP-CDFT calculation compared to MR-CDFT is used
 \beq 
 \label{eq:SP-CDFT_error}
 \sigma[O]
 =\sqrt{\frac{1}{N_i}\sum_{i=1}^{N_i} \Bigg(O^{\rm SP-CDFT}_i - O^{\rm MR-CDFT}_i \Bigg)^2}.
 \eeq 
Table~\ref{tab:Emulator_errors} presents the standard deviations  and their relative deviations  for these four quantities in \nuclide[150]{Nd} and \nuclide[150]{Sm}. The ground-state energy $E(0^+_1)$ and the proton radius $R_p$ are reproduced with relative errors of 0.02\% and 0.2\%, respectively. However, the emulator error for the excitation energy $E_x(2^+_1)$, which is several orders of magnitude smaller than the binding energy, is relatively larger, with a relative error of 13\%. The $E2$ transition strengths are reproduced with better accuracy. We also checked the relative errors of the SP-CDFT calculations for \nuclide[136]{Xe} and \nuclide[136]{Ba}, finding them to be generally less than 6\%.

\begin{table}
\tabcolsep=3pt
\caption{The standard deviations $\sigma(O)$ and relative deviations $\cal{R}(O)=\sigma(O)/O^{\rm MR-CDFT}$ of the SP-CDFT  calculations,  compared to the results of MR-CDFT, based on 64 parameter sets of EDFs for the ground-state energy $E(0^+_1)$  and proton radius  $R_p$, the excitation energy $E_x(2^+_1)$, and $B(E2: 0^+_1 \to 2^+_1)$  of \nuclide[150]{Nd} and \nuclide[150]{Sm}. } 
\renewcommand{\arraystretch}{1.4}  
\begin{tabular}{ccccccccc}
\hline 
        & \multicolumn{2}{c}{$ E(0^+_1)$ (MeV)}          &   \multicolumn{2}{c}{$R_p$ (fm)} & \multicolumn{2}{c}{$E_x(2^+_1)$ (MeV)} & \multicolumn{2}{c}{$B(E2)$ ($e^2$b$^2$)}  \\ 
 Nuclei & $\sigma$   & $\cal{R}[\%]$ & $\sigma$  & $\cal{R}[\%]$ & $\sigma$  & $\cal{R}$  & $\sigma$  & $\cal{R}[\%]$ \\
\hline 
$^{150}$Nd  &0.272&0.02 &0.003& 0.2&0.006& 4 &0.063&  2\\  
$^{150}$Sm  &0.180&0.01 &0.005& 0.1 & 0.040& 13 &0.071&  4\\
\hline
\hline
\end{tabular}
\label{tab:Emulator_errors}
\end{table}


\begin{table} 
\tabcolsep=3pt
\caption{The values of the PC-PK1 parameter set of the relativistic EDF and the ranges of the parameters in the training and target sets of the EDF.} 
\begin{tabular}{ ccccc }
\hline
\hline
 $c_\ell$ & PC-PK1~\cite{Zhao:2010PRC} & dimension & training sets[\%]  &  target sets[\%] \\  
 \hline
$\alpha_S$ & $-$3.96291$\times 10^{-4}$ & MeV$^{-2}$ &0.5&0.1\\

$\beta_S$ & +8.6653$\times 10^{-11}$ &MeV$^{-5}$  & 2 & 1\\

$\gamma_S$ &  $-$3.80724$\times 10^{-17}$&MeV$^{-8}$   & 4 &2\\

$\delta_S$ &  $-$1.09108$\times 10^{-10}$&MeV$^{-4}$  & 20 &4\\

$\alpha_V$ & +2.69040$\times 10^{-4}$ &MeV$^{-2}$  & 0.8&0.2 \\

$\gamma_V$ & $-$3.64219$\times 10^{-18}$& MeV$^{-8}$ & 30 &5\\

$\delta_V$ &  $-$4.32619$\times 10^{-10}$ &MeV$^{-4}$  & 20 &5\\

$\alpha_{TV}$ &  +2.95018$\times 10^{-5}$& MeV$^{-2}$ & 10 &6\\

$\delta_{TV}$ & $-$4.11112$\times 10^{-10}$& MeV$^{-4}$ & 150 &40\\
\hline
\hline
\end{tabular}
\label{tab:sampling_ranges}
\end{table}

\begin{figure}
    \centering
    \includegraphics[width=0.9\columnwidth]{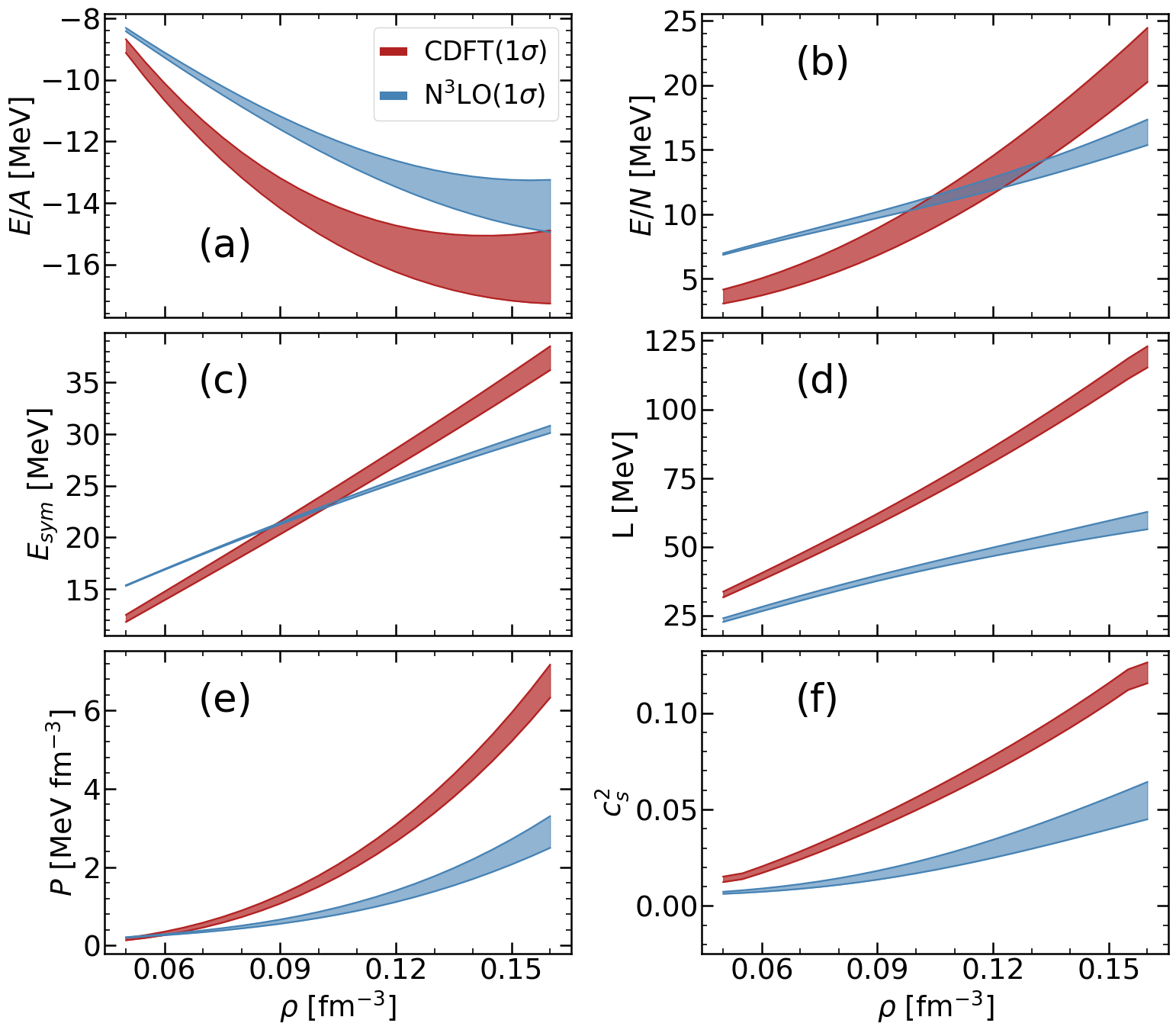}
     \caption{(Color online) (a) The energy per particle $E/A$ for symmetric nuclear matter, (b) the energy per neutron $E/N$ for pure neutron matter, (c)  symmetry energy $E_{\rm sym}$,  (d) the symmetry energy slope $L$, (e) pressure $P$, and (f) the square of the speed of sound $c^2_s$ are calculated using about one million samples from the covariant EDF. The results are compared with those, labeled as N$^3$LO, from ab initio many-body perturbation theory calculations based on a chiral Hamiltonian by Drischler et al.~\cite{Drischler:2020PRL}.}
   \label{fig:nuclear_matter_CDFT}
\end{figure}

\begin{figure}
    \centering
      \includegraphics[width=0.9\columnwidth]{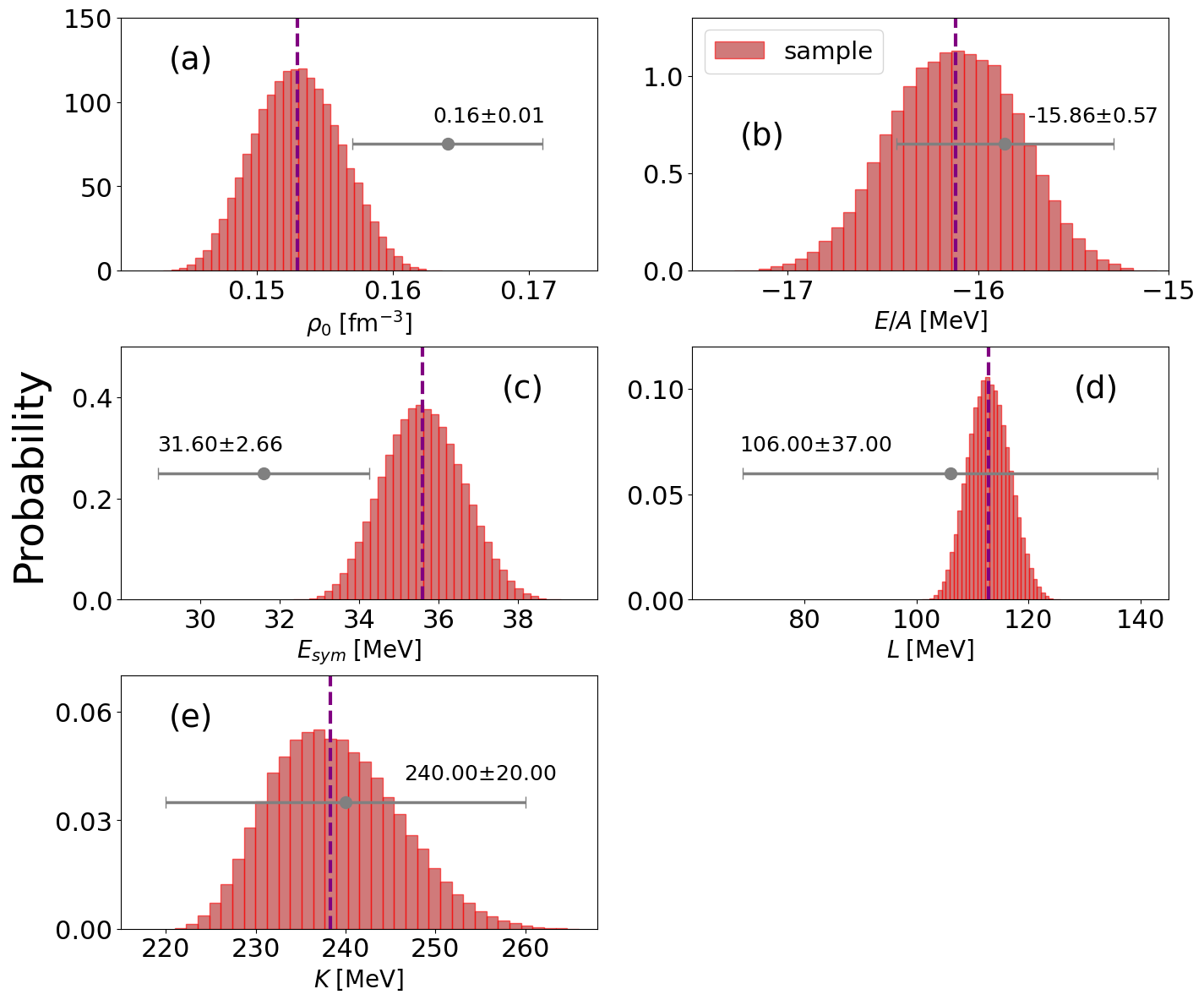}
     \caption{(Color online) Histogram plots of nuclear-matter properties around saturation density calculated using $(9+1)\times 2^{17}= 1310720$ quasi MC samples of the nine coupling constants in the relativistic EDFs around the PC-PK1 parametrization~\cite{Zhao:2010PRC}. The empirical values  (gray error bar) are given for comparison. The numbers are empirical values.}
   \label{fig:nuclear_matter_sat}
\end{figure}

\subsection{Symmetric nuclear matter}

The single-particle state in infinite nuclear matter is labeled with momentum $\mathbf{p}=\hbar\mathbf{k}$ and spin direction $\lambda$. The corresponding single-particle wave function is simplified into $u_\tau(\mathbf{k},\lambda)$, which is determined by the following Dirac equation 
    \begin{equation}
    \label{eq:nmdeq}
        \left(\bm{\alpha}\cdot\mathbf{k} + \beta M^*_\tau
         +\Sigma_{\tau}^{0}\right)u_\tau(\mathbf{k},\lambda) 
        = E_\tau(\mathbf{k}) u_\tau(\mathbf{k},\lambda),
    \end{equation}
    where $\tau(n,~p)$ distinguishes neutrons and protons, 
    $M^*_\tau = M_\tau +  \Sigma_{\tau S}$ is the Dirac mass. 
    The nucleon self-energies  are determined by the scalar and vector densities
    \begin{subequations}\label{eq:nmsnp}
    \begin{align}
        &\Sigma_{\tau S} = \alpha_S\rho_S+\beta_S\rho_S^2
        +\gamma_S \rho_S^3,
            \\
        &\Sigma_{\tau}^{0}=\alpha_V\rho_V +\gamma_V\rho^3_V
        + \alpha_{TV}\tau_3\rho_{TV},
    \end{align}
    \end{subequations}
    where $\rho_i = \rho^{(n)}_i + \rho^{(p)}_i$ with $i=S, V$ labeling scalar and vector densities, respectively. The isovector density is defined as the difference between the neutron and proton number densities, $\rho_{TV} = \rho^{(n)}_V-\rho^{(p)}_V$.  All the densities are constant for uniformly distributed nuclear matter, for which, all the derivative terms depending on the parameters $\delta_{S, V, TV}$ vanish. For neutrons ($\tau=n$) and protons ($\tau=p$), the scalar ($\rho_S$) and time-like component ($\rho_V$) of the vector densities $j^\mu_V$ are calculated as follows,
    \begin{subequations}
    \begin{align}
        \rho^{(\tau)}_{S}&=\sum_\lambda\int^{k_{\tau F}}
        \frac{d^3k}{(2\pi)^3} \frac{M^*_\tau}
        {E_\tau^*(\mathbf{k})}  ,
            \\
        \rho^{(\tau)}_{V} &=\sum_\lambda\int^{k_{\tau F}}
        \frac{d^3k}{(2\pi)^3}=\frac{k_{\tau F}^3}
        {3\pi^2},
    \end{align}
    \end{subequations} 
     where $E_{\tau F}^* = \sqrt{M^{*2}_\tau + k_{\tau F}^2}$, and $\lambda$ runs over spin up and down, leading to a factor of two. At the zero temperature, the energy density and pressure of nuclear  matter are given by~\cite{Dutra:2014PRC}
    \begin{subequations}
    \begin{align}\nonumber 
        \varepsilon=&\sum_\tau\left[\frac{3}{4}\left(
        \rho^{(\tau)}_{V} E_{\tau F}^*-\rho^{(\tau)}_{S}
        M^*_\tau\right)+\rho^{(\tau)}_{V}  M_\tau \right]
                \\  
            &+\frac{\alpha_S}{2}\rho_S^2+\frac{\beta_S}{3}\rho_S^3 + 
        \frac{\gamma_S}{4}\rho_S^4 +\frac{\alpha_V}{2}\rho^2_V  
         +\frac{\gamma_V}{4}\rho^4_V+ \frac{\alpha_{TV}}{2}\rho_{TV}^2,
                \\  \nonumber 
        p =&\sum_\tau\left[\frac{1}{4}\left(\rho^{(\tau)}_{V}E_{\tau F}^*
            -\rho^{(\tau)}_{ S}M_\tau^*\right)+\Sigma_{\tau 0}\rho^{(\tau)}_{V} 
            +\Sigma_{\tau S}
		\rho^{(\tau)}_{S}\right]
                \\
	   &-\frac{\alpha_S}{2}\rho_S^2-\frac{\beta_S}{3}\rho_S^3- 
		\frac{\gamma_S}{4}\rho_S^4-  \frac{\alpha_V}{2}\rho^2_V-  \frac{\gamma_V}{4}\rho^4_V
  - \frac{\alpha_{TV}}{2}\rho_{TV}^2. 
    \end{align}
    \end{subequations} 
 The average nucleon binding energy $E/A$ and incomprehensibility $K$ are defined as 
 \beq 
 E/A \equiv \varepsilon/\rho_V -M,\quad K \equiv 9\frac{\partial p}{\partial \rho_V}.
 \eeq 
 Besides, the symmetry energy $E_\mathrm{sym}$ and its slope parameter $L$ are calculated by
    \begin{equation}\label{eq:nmesym}
        E_\mathrm{sym} \equiv \frac{1}{2}\left.\frac{\partial^2 (E/A)}
        {\partial \eta^2}\right|_{\eta=0},\qquad 
        L \equiv 3\rho
        \frac{\partial E_\mathrm{sym}}{\partial \rho_V},
    \end{equation}
    with the isospin asymmetry $\eta = \rho_{TV}/\rho_V$.  The speed of sound is usually defined as $c_s=\sqrt{\partial p/\partial \epsilon}$~\cite{Drischler:2020PRL}, in units of the speed of light $c$.

We sampled a total of $(9+1) \times 2^{17} \approx 1.3 \times 10^6$ EDF parameter sets by varying all nine parameters around the PC-PK1 values~\cite{Zhao:2010PRC} using quasi Monte-Carlo (MC) sampling with a uniform distribution within the ranges shown in Table~\ref{tab:sampling_ranges}. Using these parameter sets, we calculated the properties of infinite nuclear matter as a function of nucleon number density $\rho_V$, as illustrated in Fig.~\ref{fig:nuclear_matter_CDFT}. For comparison, results from many-body perturbation theory based on a chiral Hamiltonian~\cite{Drischler:2020PRL} are also presented, revealing significant discrepancies between the two approaches. The physical quantities at saturation density $\rho_0$, denoted as $\Theta_{\rm sat} = \{\rho_0, E/A, E_{\rm sym}, L, K\}$, calculated from the sampled EDF parameter sets, are displayed in Fig.\ref{fig:nuclear_matter_sat}. We observe some mismatches between the mean values of $\rho_0$ and $E_{\rm sym}$ and their empirical values. To incorporate refinements from nuclear matter into the EDFs, we introduce the implausibility function \cite{Vernon:2010}
\begin{equation}
\begin{aligned}
    I_{(i)}(\mathbf{C})
    &=\sqrt{\frac{\left[O^{\rm calc.}_{(i)}(\mathbf{C})-O^{\rm empi.}_{(i)}\right]^{2}}{ \sigma^2\left(O^{\rm empi.}_{(i)}\right)}}.
\end{aligned}
\label{hm}
\end{equation}
Here, $\sigma\left(O^{\rm empi.}{(i)}\right)$ denotes the standard deviation of the empirical value for the $i$-th physical quantity of infinite nuclear matter. The implausibility function $I{(i)}(x)$ measures the probability that the theoretically calculated value $O^{\rm calc.}{(i)}(x)$ matches the empirical value $O^{\rm empi.}{(i)}$. We screen the parameter sets $\mathbf{C}$ based on the criterion $\max[I_{(i)}(x)] < \sqrt{3}$, which corresponds to a probability of about 92\%. Here, $\max[I_{(i)}(x)]$ represents the largest value of the implausibility functions across the physical quantities $\Theta_{\rm sat}$ of infinite nuclear matter. After applying the $\sqrt{3}\sigma$ screening rule using the properties of nuclear matter, we obtain a final set of 457,380 non-implausible samples.  

\subsection{Nuclear low-lying states}

\begin{figure*}[]
    \centering 
    \includegraphics[width=2\columnwidth]{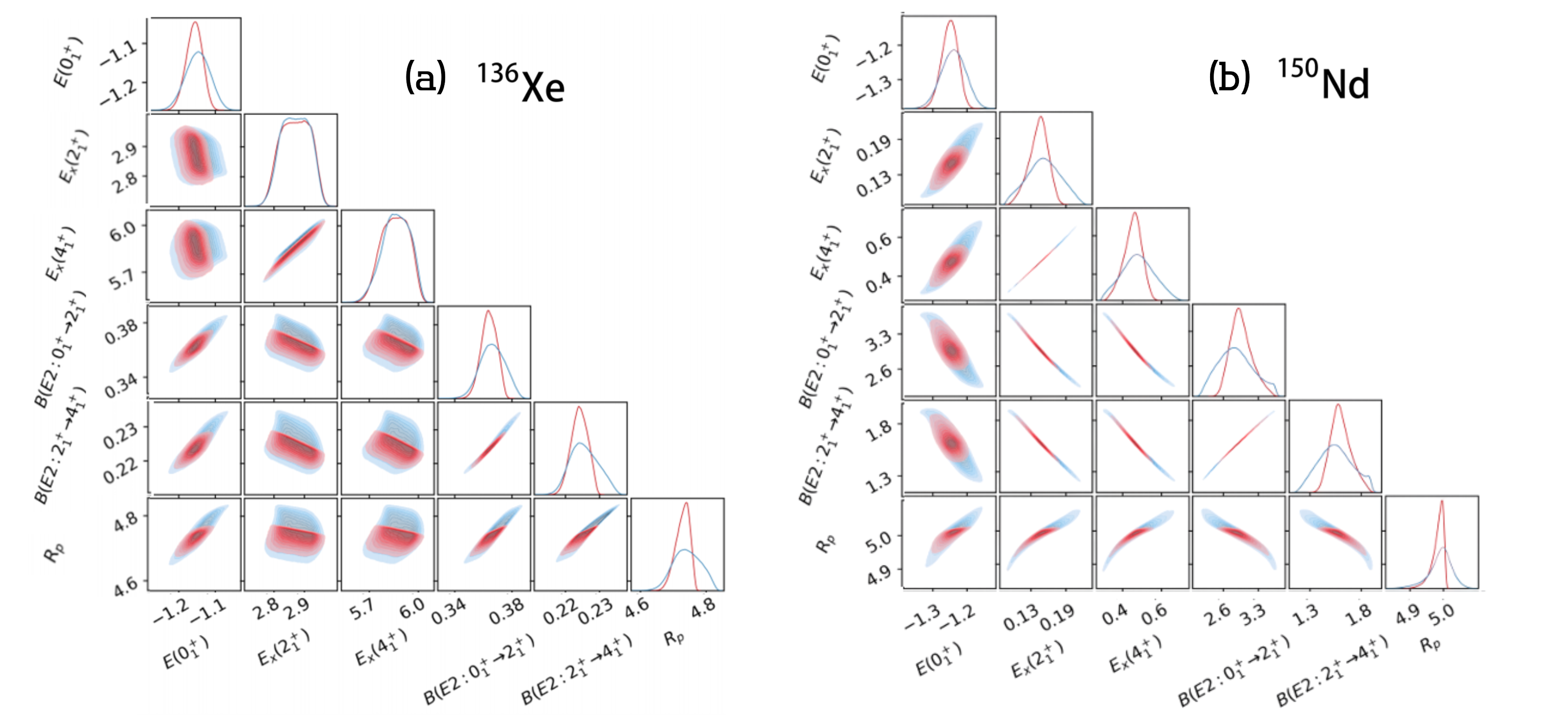}  
     \caption{(Color online) The correlation relation among different quantities of low-lying states. The diagonal diagrams  are the histograms for the probability distributions  of these quantities  for (a) \nuclide[136]{Xe} and (b) \nuclide[150]{Nd}. The red distributions are refined by the empirical values of the physical quantities $\Theta_{\rm sat}=\{\rho_0, E/A, E_{\rm sym}, L, K\}$ of infinite nuclear matter at saturation density using $\sqrt{3}\sigma$ rule.  See main text for details. }
   \label{fig:correlation4Xe_Nd}
\end{figure*}

\begin{figure}[]
    \centering  
        \includegraphics[width=\columnwidth]{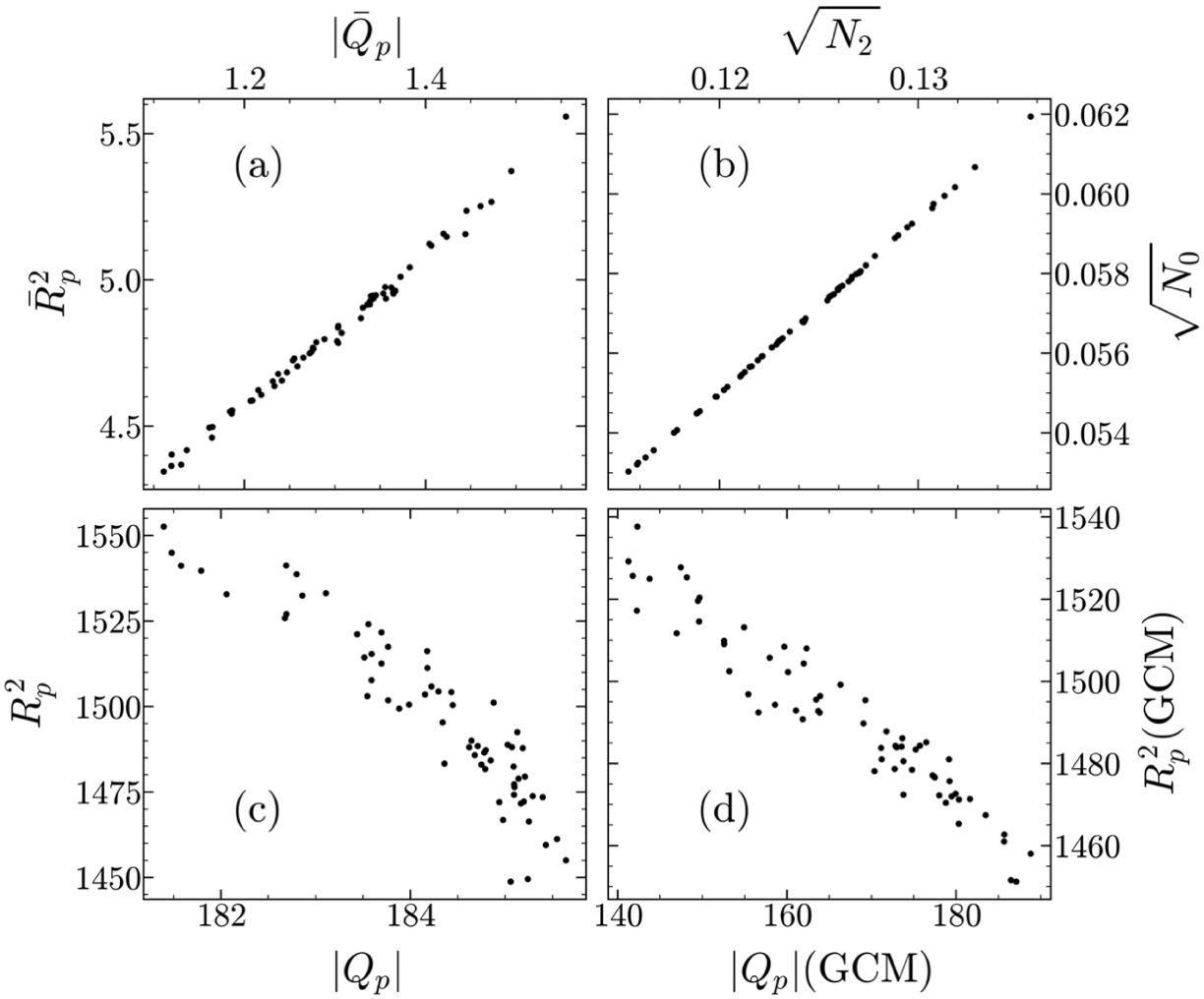}  
     \caption{(a), (b), (c) The correlations  among mean-squared radius $R_p^2=N^{-1}_0\bra{J=0NZ;\beta_2,\mathbf{C}}\hat R^2_p\ket{J=0NZ;\beta_2,\mathbf{C}}$, reduced $E2$ transition matrix element $Q_p=(N_0N_2)^{-1/2}\bra{J=2NZ;\beta_2,\mathbf{C}}|\hat Q_2|\ket{J=0NZ;\beta_2,\mathbf{C}}$ and normalization factors $\sqrt{N_{J}}$ from the calculations for \nuclide[150]{Nd} based on a single configuration with $\beta_2=0.3$, where $ N_J=\bra{JNZ;\beta_2,\mathbf{C}}JNZ;\beta_2,\mathbf{C}\rangle$. (d) The results from the configuration-mixing GCM calculations. The quantities hatted with bars are unnormalized. See main text for details.}
   \label{fig:correlation_analysis_Rp_BE2}
\end{figure}

Using the above non-implausible samples, we calculated the physical quantities of nuclear low-lying states with SP-CDFT(14, $k_{\rm max}$) for \nuclide[150]{Nd}, \nuclide[150]{Sm}, \nuclide[136]{Xe}, and \nuclide[136]{Ba}. The results for \nuclide[136]{Xe} and \nuclide[150]{Nd}, with and without refinement from nuclear matter properties, are shown in blue and red in Fig.~\ref{fig:correlation4Xe_Nd}. Since the results for \nuclide[150]{Sm} closely resemble those for \nuclide[150]{Nd}, and those for \nuclide[136]{Ba} are similar to \nuclide[136]{Xe}, we do not display them here.  It is shown that  the excitation energies of the $2^+_1$ and $4^+_1$ states are anti-correlated with the $E2$ transition strengths. This correlation is strong in the deformed nuclei \nuclide[150]{Nd} and \nuclide[150]{Sm}, but very weak in the spherical nuclei \nuclide[136]{Xe} and \nuclide[136]{Ba}. Interestingly, the proton radius $R_p$ of the ground state is positively correlated with the $B(E2: 0^+_1 \to 2^+_1)$ in \nuclide[136]{Xe}, while it is anti-correlated in \nuclide[150]{Nd}. To understand the anti-correlation behavior in \nuclide[150]{Nd}, we plot the unnormalized mean-squared radii $\bar{R}_p^2$ against the unnormalized reduced $E2$ transition matrix element $|\bar{Q}_p|$ from single-configuration calculations using different EDF parameter sets in Fig.\ref{fig:correlation_analysis_Rp_BE2}. These are shown to be strongly positively correlated, as expected. Since the normalization factor $\sqrt{N_0}$ of the $J=0$ state is also linearly correlated with the normalization factor $\sqrt{N_2}$ of the $J=2$ state with a nonzero intercept, it results in an anti-correlation between the normalized mean-squared radii $R_p^2$ and the normalized reduced $E2$ transition matrix element $|Q_p|$, as illustrated in Fig.\ref{fig:correlation_analysis_Rp_BE2}(c). This anti-correlation persists in the configuration-mixing GCM calculation, as shown in Fig.~\ref{fig:correlation_analysis_Rp_BE2}(d). In short, the $B(E2: 0^+_1 \to 2^+_1)$ could be either positive or negative correlated with the proton radius $R_p$ of the ground state.

\begin{figure}[]
    \centering
    \includegraphics[width=\columnwidth]{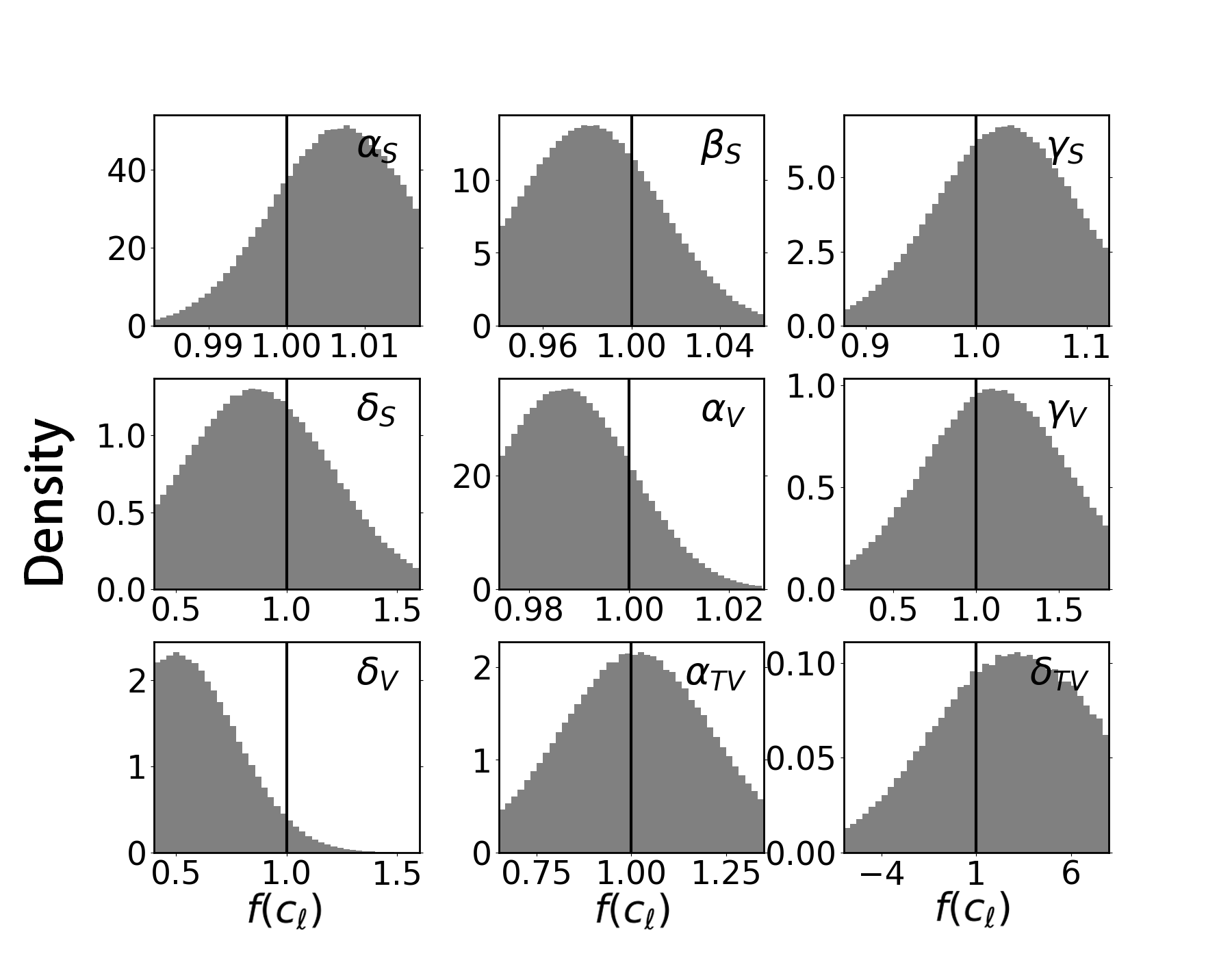}
    \caption{The posterior distributions $p(\mathbf{C}|\mathcal{D})$, defined in \eqref{eq:post_parameter_data}, for the nine parameters (normalized to PC-PK1) in the relativistic EDF from the Bayesian analysis, where the data $\mathcal{D}$ contain the empirical values of nuclear matter at saturation density $\mathbf{\Theta}_{\rm sat}$, the $\mathbf{\Theta}_{\rm low}$ of nuclear matter below saturation density from the chiral nuclear force,  and the data for the $B(E2:0_1^+\to 2_1^+)$ of $^{136}$Xe.  }
    \label{fig:post_parameters_NM_Xe136}
\end{figure}
 
\begin{figure}[]
    \centering
    \includegraphics[width=\columnwidth]{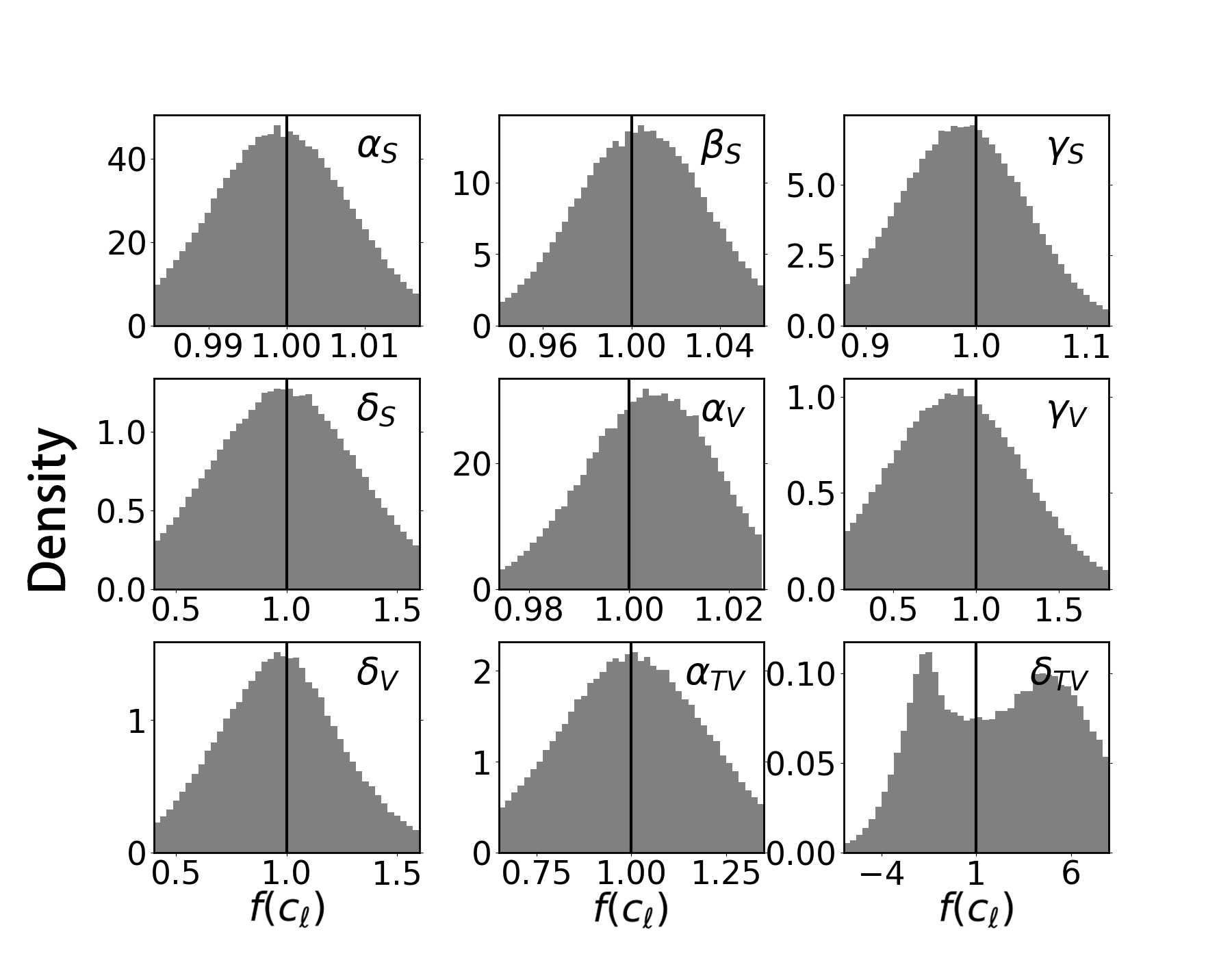}
    \caption{Same as Fig.~\ref{fig:post_parameters_NM_Xe136}, but replacing the data for the $B(E2:0_1^+\to 2_1^+)$ of $^{136}$Xe with that of \nuclide[150]{Nd}.}
    \label{fig:post_parameters_NM_Nd150}
\end{figure}

Subsequently, we use the Bayesian method to derive the posterior distribution $p(\mathcal{O}|\mathcal{D})$ for the quantity $\mathcal{O}$ given the data $\mathcal{D}$. The true value $\mathcal{O}^{(\rm{true})}$ for the quantity $\mathcal{O}$ can be decomposed as follows \cite{Wesolowski:2019},
\begin{equation}
\mathcal{O}^{(\rm{true})}
=\mathcal{O}^{(\rm{MR-CDFT})}(\mathbf{C}) +\delta^{(\rm{syst})}
\end{equation}
where the $\mathcal{O}^{(\rm{MR-CDFT})}(\mathbf{C})$ represents the value by the MR-CDFT based on the parameter set $\mathbf{C}$, and it can be further divided into two terms in this work,
\begin{equation}
\mathcal{O}^{(\rm{MR-CDFT})} (\mathbf{C})=\mathcal{O}^{(\rm{SP-CDFT})}(\mathbf{C})+\delta^{(\rm{em})}.
\end{equation}
Both the  systematic error and the emulator error are assumed to follow normal distributions with zero mean,
\begin{equation}
\delta^{\mathrm{(em)}} \sim \mathcal{N}(0, \sigma_{(\mathrm{em})}), \quad
\delta^{\mathrm{(syst)}} \sim \mathcal{N}(0, \sigma_{\mathrm{(syst)}}),
\end{equation}
and are treated as independent. The emulator uncertainty $\sigma_{(\mathrm{em})}$ is estimated from benchmark comparisons between SP-CDFT and MR-CDFT calculations in Eq.(\ref{eq:SP-CDFT_error}), while $\sigma_{(\mathrm{syst})}$ is determined from the typical deviation between EDF predictions with PC-PK1 and empirical data for the quantities considered.

This posterior distribution can be expressed as an integral that incorporates statistical information from various parameter sets:
\begin{equation}
   \label{eq:posterior_O_Data}
p(\mathcal{O}|\mathcal{D})=\int p(\mathcal{O}|\mathbf{C}) p(\mathbf{C}|\mathcal{D}) d\mathbf{C},
\end{equation}
where $p(\mathbf{C}|\mathcal{D})$ represents the posterior distribution of the model parameters $\mathbf{C}$ and can be determined using Bayes' theorem:
\begin{equation}
\label{eq:post_parameter_data} 
p(\mathbf{C}|\mathcal{D})=\frac{p(\mathcal{D}|\mathbf{C}) \pi(\mathbf{C})}{p(\mathcal{D})}.
\end{equation}
In this equation, $p(\mathcal{D})$ serves as a normalization constant.  The prior distribution $\pi(\mathbf{C})$ of the parameters reflects the empirical evaluation of each parameter set $\mathbf{C}$. Here, we use an uncorrelated multivariate normal distribution,
\beq
\pi({\rm C}) \propto  \exp(-\chi_0^2/2),\quad \chi_0^2=\sum_{\ell=1}^{9}\frac{(c_\ell-c_\ell^0)^2}{\sigma_\ell^2},
\eeq 
where $c_\ell^0$ is the value of the $\ell$-th parameter in the PC-PK1 set, and $\sigma_\ell$ is the standard deviation of the parameter $c_\ell$ in the samples from the quasi MC sampling method as mentioned before.

\begin{table}
\tabcolsep=3pt
\renewcommand{\arraystretch}{1.5}
\caption{The median, 4th percentile, and 96th percentile (corresponding to 92\% C.L.) of the excitation energies  (in MeV) of $2^+_1$ and $4^+_1$ states and $E2$ transition strengths (in e$^2b^2$) derived from the  posteriors of the SP-CDFT calculations for  \nuclide[150]{Nd}, \nuclide[150]{Sm}, \nuclide[136]{Xe}, and \nuclide[136]{Ba}, in comparison with available data~\cite{NNDC}.}
\begin{tabular}{lllll}
\hline
\hline
 & $E_x(2^+_1)$ & $E_x(4^+_1)$ & $B(E2:2^+_1\to 4^+_1)$ & $B(E2:0^+_1\to 2^+_1)$\\
\hline
$^{150}$Nd \\ 
Exp. &0.130 &0.381 &1.539(14) &2.745(70)\\
Calc.  &$0.146^{+0.022}_{-0.027}$& $0.461^{+0.057}_{-0.068}$&$1.579^{+0.131}_{-0.101}$&$2.918^{+0.309}_{-0.250}$\\ 
\hline
$^{150}$Sm \\
Exp.  &0.334 & 0.773&0.937(145) &1.351(31)\\
Calc.  &$0.301^{+0.041}_{-0.062}$& $0.789^{+0.059}_{-0.094}$&$1.112^{+0.081}_{-0.054}$ &$1.913^{+0.226}_{-0.134}$\\ 
\hline
$^{136}$Xe \\
Exp.  &1.313 & 1.694&0.011  &0.230(9) \\
Calc.  &$2.872^{+0.069}_{-0.072}$& $5.864^{+0.117}_{-0.129}$&$0.224^{+0.004}_{-0.004}$ &$0.364^{+0.008}_{-0.009}$\\ 
\hline
$^{136}$Ba \\
Exp.  &0.819 & 1.551& 0.119(51)  &0.464(9)\\
Calc.  &$1.756^{+0.065}_{-0.086}$& $3.548^{+0.195}_{-0.200}$&$0.278^{+0.019}_{-0.018}$ &$0.451^{+0.020}_{-0.021}$ \\
\hline\hline
\end{tabular}
\label{tab:UQ_low-lying-states}
\end{table}

The likelihood function $p(\mathcal{D}|\mathbf{C})$ also takes the form of multivariate normal distribution 
\beqn 
\label{eq:likelihood}
 && p( \mathcal{D}|\mathbf{C}) \nonumber\\ &\propto & \exp\left[-\frac{1}{2}\Bigg(\mathbf{D}^{\rm (em)}({\rm \mathbf{C}})-\mathbf{D}^{\rm (exp)}\Bigg)^T \Sigma^{-1}\Bigg(\mathbf{D}^{\rm (em)}({\rm \mathbf{C}})-\mathbf{D}^{\rm (exp)}\Bigg)\right], \notag\\ 
 \eeqn
 where $\mathbf{D}^{\rm (em)}({\rm \mathbf{C}})$ denote the values of a set of quantities predicted by the emulator based on the parameter set ${\rm \mathbf{C}}$, and $\mathbf{D}^{\rm (exp)}$ are either the corresponding data or empirical values. The obtained $p(\mathcal{D}|\mathbf{C})$ of the quantities $\mathcal{D}$, including nuclear matter and $B(E2)$, is used to constrain the parameter ${\rm \mathbf{C}}$ with the Bayesian method.

 The covariance matrix $\Sigma_{ij}$ is obtained from the Pearson coefficient $\rho_{ij}$ by $\Sigma_{ij}=\sigma_i\rho_{ij} \sigma_j$, where the $\sigma_{i}$ denotes the standard deviation of the SP-CDFT calculation for the $i$-th quantity, in comparison with the corresponding data or empirical value.  The  Pearson coefficient  $\rho_{ij}$ is defined by the expectation values  
\beq
\rho_{ij}=
\frac{\mathbb{E}\big[(D_i-\mu_i)(D_j-\mu_j)\big]}
{\sqrt{\mathbb{E}[(D_i-\mu_i)^2]}\,\sqrt{\mathbb{E}[(D_j-\mu_j)^2]}}
\eeq
where $\mu_i$ is the mean value of the $i$-th quantity. The covariance matrix encapsulates the correlation between $i$-th and $j$-th quantities.

The predictive distribution $ p(\mathcal{O}|\mathbf{C})$ in Eq.\eqref{eq:posterior_O_Data} is  given by
\beqn    
p(\mathcal{O}|\mathbf{C})=\int p(\mathcal{O}|\mathbf{C};\sigma_{\rm{(syst)}})p(\sigma_{\rm{(syst)}})d\sigma_{\rm{(syst)}},
\eeqn
with
\beqn 
\label{eq:emulator}   
p(\mathcal{O}|\mathbf{C};\sigma_{\rm{(syst)}})
&\propto&\exp\Bigg\{-\frac{[\mathcal{O}-\mathcal{O}^{\rm (em)}(\mathbf{C})]^2}{2(\sigma_{(\rm{em})}^2+\sigma_{(\rm{syst})}^2)}\Bigg\},
\eeqn
where  $\mathcal{O}^{\rm (em)}(\mathbf{C})$ is the prediction of the emulator for the  quantity $\mathcal{O}$ using the parameter set $\mathbf{C}$, and $\sigma_{\rm{(em)}}$ is the emulator error of this quantity. In this work, the prior of the systematic error $p(\sigma_{\rm{(syst)}})$ is chosen as a delta function for observable with experimental data, i.e., $p(\mathcal{O}|\mathbf{C})=p(\mathcal{O}|\mathbf{C};\sigma_{\rm{(syst)}})$.

Figure~\ref{fig:post_parameters_NM_Xe136} shows the posterior distribution  $p(\mathbf{C}|\mathcal{D})$ of each parameter $c_\ell$, where the data $\mathcal{D}$ contain the empirical values of the nuclear-matter properties $\mathbf{\Theta}_{\rm sat}$ at saturation density, those below saturation density $\mathbf{\Theta}_{\rm low}$ obtained from the many-body perturbation theory calculation based on a chiral $NN+3N$ potential up to N$^3$LO~\cite{Drischler:2020PRL}, and the data for the $B(E2:0_1^+\to 2_1^+)$ of $^{136}$Xe. Figure~\ref{fig:post_parameters_NM_Nd150} shows the posterior distribution  $p(\mathbf{C}|\mathcal{D})$ which is obtained similar to Fig.~\ref{fig:post_parameters_NM_Xe136}, but replacing the $B(E2:0_1^+\to 2_1^+)$ of $^{136}$Xe with that of $^{150}$Nd. It is seen that the main peaks of the posterior distribution $p(\mathbf{C}|\mathcal{D})$ for most parameters derived from the $B(E2)$ data of $^{136}$Xe are offset from their values in the PC-PK1. In contrast, the parameters derived from the $B(E2)$ data of $^{150}$Nd align well with the values in the PC-PK1. This phenomenon can be understood from the observation that  the $B(E2)$ of $^{150}$Nd is much better described with the present MR-CDFT than that for $^{136}$Xe, as detailed in  Table~\ref {tab:UQ_low-lying-states}. It lists the median and uncertainties of the posteriors for the excitation energies of the $2^+_1$ and $4^+_1$ states and $E2$ transition strengths in the four nuclei. It is clear that the excitation energies of the deformed nuclei \nuclide[150]{Nd} and \nuclide[150]{Sm} are excellently reproduced, whereas those of the near-spherical nuclei \nuclide[136]{Xe} and \nuclide[136]{Ba} are overestimated. The ratio $R_{42}=E_x(4^+_1)/E_x(2^+_1)<2$ in both \nuclide[136]{Xe} and \nuclide[136]{Ba} indicates that their $2^+_1, 4^+_1$ states are dominated by the seniority coupling~\cite{Qi:2017PLB}, the description of which requires the inclusion of  non-collective excitation configurations~\cite{Yao:2015}.  The weak collective nature of the $2^+_1, 4^+_1$ states in \nuclide[136]{Xe} and \nuclide[136]{Ba} can also be inferred from the weak $E2$ transition strengths. With the posterior distribution, we finally derive the statistical uncertainties for the low-lying states, which are generally within 21\% for excitation energies and  12\% for $E2$ transition strengths.

 \section{Summary}
 \label{sec:summary}
In this work, we have presented a comprehensive formulation of the subspace-projected covariant density functional theory (SP-CDFT), a novel framework that combines the eigenvector continuation (EC) method with the quantum-number projected generator coordinate method (PGCM) within covariant density functional theory (CDFT). We demonstrate that SP-CDFT provides an efficient and accurate emulator of multireference (MR)-CDFT for the description of nuclear low-lying states. The emulator errors are found to be within a few tenths of a percent for bulk properties, and at the level of a few percent for excitation energies and $E2$ transition strengths.

Building on this framework, we quantify statistical uncertainties in both nuclear-matter properties and nuclear low-lying states for \nuclide[136]{Xe}, \nuclide[136]{Ba}, \nuclide[150]{Nd}, and \nuclide[150]{Sm}, based on the PC-PK1 EDF within a Bayesian approach. Constraints from nuclear-matter properties are employed to refine the parameter sets used in calculations for finite nuclei. The analyzed observables include ground-state energies, proton radii, excitation energies of the $2^+_1$ and $4^+_1$ states, and $E2$ transition strengths. We find that the $B(E2: 0^+_1 \to 2^+_1)$ values can exhibit either positive or negative correlations with the ground-state proton radius $R_p$, depending on the nuclear structure. Furthermore, the propagated statistical uncertainties associated with the nine EDF parameters reach up to $21\%$ for excitation energies and $12\%$ for $E2$ transition strengths.

These results, together with the comparison between theoretical predictions and experimental data, highlight the intrinsic limitations of the underlying EDF framework. After accounting for statistical uncertainties, the excitation energies and $B(E2)$ values of deformed nuclei are well reproduced, whereas those of near-spherical nuclei remain challenging. Future work will focus on improving the description of near-spherical systems within SP-CDFT and on further constraining nuclear EDFs using data on low-lying nuclear states.

\section*{Acknowledgments} 
We thank K. Hagino, H. Hergert, W.G. Jiang, C.F. Jiao, X.L. Zhang, and Y.N. Zhang for valuable discussions. We are grateful to C. Drischler for providing nuclear-matter properties from \textit{ab initio} calculations. This work was supported in part by the National Natural Science Foundation of China under Grant Nos. 125B2108, 12405143, and 12375119. We also acknowledge the Beijing Super Cloud Computing Center (BSCC) for providing high-performance computing resources.

 
%

\end{document}